\documentclass[times,twocolumn,final]{elsarticle}

\usepackage{framed,multirow}

\usepackage{amssymb}
\usepackage{latexsym}
\usepackage{lineno}

\usepackage{url}
\usepackage[dvipsnames]{xcolor}
\usepackage{hyperref}

\usepackage{amsmath,amssymb,amsfonts}
\usepackage{breqn}

\usepackage{algorithm}
\usepackage{algpseudocode}

\usepackage{graphicx}
\usepackage{textcomp}
\usepackage{thmtools, thm-restate}

\usepackage{epsfig}
\usepackage{tabularx}
\usepackage{booktabs}
\usepackage{verbatim}
\usepackage{xcolor}
\usepackage{amstext}
\usepackage{upgreek}
\usepackage{multirow}
\usepackage{bbm}
\usepackage{rotating}
\usepackage{pifont}
\usepackage{colortbl}

\newcommand{\ttt}{\boldsymbol{\theta}}
\newcommand{\ppp}{\boldsymbol{\phi}}

\newcommand{\xx}{\mathbf{x}}
\newcommand{\aaa}{\mathbf{a}}
\newcommand{\mm}{\mathbf{m}}

\newcommand{\zz}{\mathbf{z}}
\newcommand{\real}{\mathbb{R}}

\usepackage{subfig}

\usepackage{amsfonts}

\definecolor{mypink1}{rgb}{0.858, 0.188, 0.478}
\definecolor{myorange}{RGB}{255,165,0}
\definecolor{gray}{cmyk}{0.86,0.86,0.86,0.86}

\definecolor{newcolor}{rgb}{.8,.349,.1}

\journal{Medical Image Analysis}

\begin{document}


\begin{frontmatter}


\title{Constrained unsupervised anomaly segmentation}

\author[1]{Julio Silva-Rodríguez}
\cortext[cor1]{Corresponding author: jjsilva@upv.es}
\author[2]{Valery Naranjo}
\author[3]{Jose Dolz}

\address[1]{Institute of Transport and Territory, Universitat Politècnica de València, Valencia, Spain}
\address[2]{Institute of Research and Innovation in Bioengineering, Universitat Politècnica de València, Valencia, Spain}
\address[3]{\'{E}cole de Technologie Sup\'{e}rieure, Montreal, QC H3C 1K3, Canada}


\begin{abstract}
Current unsupervised anomaly localization approaches rely on generative models to learn the distribution of normal images, which is later used to identify potential anomalous regions derived from errors on the reconstructed images. To address the limitations of residual-based anomaly localization, very recent literature has focused on attention maps, by integrating supervision on them in the form of homogenization constraints. In this work, we propose a novel formulation that addresses the problem in a more principled manner, leveraging well-known knowledge in constrained optimization. In particular, the equality constraint on the attention maps in prior work is replaced by an inequality constraint, which allows more flexibility. In addition, to address the limitations of penalty-based functions we employ an extension of the popular log-barrier methods to handle the constraint. Last, we propose an alternative regularization term that maximizes the Shannon entropy of the attention maps, reducing the amount of hyperparameters of the proposed model. Comprehensive experiments on two publicly available datasets on brain lesion segmentation demonstrate that the proposed approach substantially outperforms relevant literature, establishing new state-of-the-art results for unsupervised lesion segmentation.  
\end{abstract}
        

\end{frontmatter}


\begin{abstract}
Current unsupervised anomaly localization approaches rely on generative models to learn the distribution of normal images, which is later used to identify potential anomalous regions derived from errors on the reconstructed images. To address the limitations of residual-based anomaly localization, very recent literature has focused on attention maps, by integrating supervision on them in the form of homogenization constraints. In this work, we propose a novel formulation that addresses the problem in a more principled manner, leveraging well-known knowledge in constrained optimization. In particular, the equality constraint on the attention maps in prior work is replaced by an inequality constraint, which allows more flexibility. In addition, to address the limitations of penalty-based functions we employ an extension of the popular log-barrier methods to handle the constraint. Last, we propose an alternative regularization term that maximizes the Shannon entropy of the attention maps, reducing the amount of hyperparameters of the proposed model. Comprehensive experiments on two publicly available datasets on brain lesion segmentation demonstrate that the proposed approach substantially outperforms relevant literature, establishing new state-of-the-art results for unsupervised lesion segmentation.  
\end{abstract}

\section{Introduction}
\label{sec:intro}

Deep learning models are driving progress in a wide range of visual recognition tasks, particularly when they are trained with large amounts of annotated samples. This learning paradigm, however, carries two important limitations. First, obtaining such curated labeled datasets is a cumbersome process prone to annotator subjectivity, limiting the access to sufficient training data in practice. This problem is further magnified in the context of medical image segmentation, where labeling involves assigning a category to each image pixel or voxel.
In addition, even if annotated images are available, there exist some applications, such as brain lesion detection, where large intra-class variations are not captured during training, failing to cover the broad range of abnormalities that might be present in a scan. This results in trained models which are potentially tailored to discover lesions similar to those seen during training. Thus, considering the scarcity and the diversity of target objects in these scenarios, lesion segmentation is typically modeled as an anomaly localization task, which is trained in an unsupervised manner. In this setting, the training dataset contains only \textit{normal} images and \textit{abnormal} images are not ideally accessible during training. 

A popular strategy to tackle unsupervised anomaly segmentation is to model the distribution of normal images in the training set. To this end, generative models, such as generative adversarial networks (GANs) (\cite{Schlegl2017UnsupervisedDiscovery,Schlegl2019F-AnoGAN:Networks,Andermatt2019PathologyOrigin,Ravanbakhsh2019TrainingCrowds,Baur2020SteGANomaly:MRI,Sun2020AnDetection}) and variational auto-encoders (VAEs) (\cite{Chen2018UnsupervisedAuto-encoders,NickPawlowski2018UnsupervisedAutoencoders,Sabokrou2019AVID:Detection,Chen2020UnsupervisedPrior,Zimmerer2020Abstract:Auto-encoders}) have been widely employed. In particular, these models are trained to reconstruct their input images, which are drawn from a normal, i.e., \textit{healthy}, distribution. At inference, input images are compared to their reconstructed normal counterparts, which are recovered from the learned distribution. Then, the anomalous regions are identified from the reconstruction error.

As an alternative to these methods, a few recent works have integrated class-activation maps (CAMs) during training \cite{Venkataramanan2020AttentionImages,Liu2020TowardsAutoencoders}. In particular, \cite{Venkataramanan2020AttentionImages} leverage the generated attention maps as an additional supervision cue, enforcing the network to provide attentive regions covering the whole context in normal images. This term was formulated as an equality constraint with the form of a L$_1$ penalty over each individual pixel. Nevertheless, we found that explicitly forcing the network to produce maximum attention values across each pixel does not achieve satisfactory results in the context of brain lesion segmentation. In addition, recent literature in constrained optimization for deep neural networks suggests that simple penalties --such as the function used in \cite{Venkataramanan2020AttentionImages}-- might not be the optimal solution to constraint the output of a CNN (\cite{Kervadec2019ConstrainedExtensions}).

Based on these observations, we propose a novel formulation for unsupervised semantic segmentation of brain lesions in medical images. The key contributions of our work can be summarized as follows:

\begin{itemize}
    
    \item A novel constrained formulation for unsupervised 
    lesion segmentation, which integrates an auxiliary constrained loss to force the network to generate attention maps that cover the whole context in normal images.
    
    \item In particular, we leverage \textit{global} inequality constraints on the generated attention maps to force them to be activated around a certain target value. This contrasts with the previous work in \cite{Venkataramanan2020AttentionImages}, where \textit{local} pixel-wise equality constraints on Grad-CAMs \cite{Selvaraju2020Grad-CAM:Localization} are employed. 
    In addition, to address the limitations of penalty-based functions, we resort to an extended version of the standard log-barrier.
    
    \item Furthermore, we consider an alternative regularization term that maximizes the Shannon entropy of the attention maps, reducing the amount of hyperparameters with respect to the extended log-barrier model, while yielding at par performances.

    \item We benchmark the proposed model against a relevant body of literature on two public lesion segmentation benchmarks: BraTS and Physionet-ICH datasets. Comprehensive experiments demonstrate the superior performance of our model, establishing a new state-of-the-art for this task.
\end{itemize}

This journal version provides a substantial extension of the conference work presented in \citep{Silva-Rodriguez2021LookingSegmentation}. First, we extended the literature survey, particularly for unsupervised medical image segmentation. Then, in terms of methodology, the current version introduces several important modifications. In particular, we further investigate the role of the gradients on the attention maps derived from Grad-CAM in the task of unsupervised anomaly detection. Based on our empirical observations, we modify the formulation in \cite{Silva-Rodriguez2021LookingSegmentation} to constraint directly the activation maps without involving any gradient information. Furthermore, we propose an alternative learning objective for our constrained problem based on the Shannon entropy. More concretely, we replace our log-barrier formulation by a maximizing entropy term on the softmax activation of brain tissue pixels, which reduces the complexity in terms of hyperparameters with respect to the former model. Last, we add comprehensive experiments to empirically validate our method, including an additional dataset and extensive ablation studies on several design choices.

\section{Related Work}
\label{sec:rw}

\subsection{Unsupervised anomaly segmentation}

Unsupervised anomaly segmentation aims at identifying abnormal pixels on test images, containing, for example, lesions on medical images (\cite{Baur2020SteGANomaly:MRI,Chen2018UnsupervisedAuto-encoders}), defects in industrial images (\cite{Bergmann2019ImprovingAutoencoders,Liu2020TowardsAutoencoders,Venkataramanan2020AttentionImages}) or abnormal events in videos (\cite{Abati2019LatentDetection,Ravanbakhsh2019TrainingCrowds}). A main body of the literature has explored unsupervised deep (generative) representation learning to learn the distribution from normal data. The underlying assumption is that a model trained on normal data will not be able to reconstruct anomalous regions, and the reconstructed difference can therefore be used as an anomaly score. Under this learning paradigm, generative adversarial networks (GAN) (\cite{Goodfellow2014GenerativeNetworks}) and variational auto-encoders (VAE) (\cite{Kingma2014Auto-encodingBayes}) are typically employed. Nevertheless, even though GAN and VAE model the latent variable, the manner in which they approximate the distribution of a set of samples differs. GAN-based approaches (\cite{Schlegl2017UnsupervisedDiscovery,Schlegl2019F-AnoGAN:Networks,Andermatt2019PathologyOrigin,Ravanbakhsh2019TrainingCrowds,Baur2020SteGANomaly:MRI,Sun2020AnDetection}) approximate the distribution by optimizing a generator to map random samples from a prior distribution in the latent space into data points that a trained discriminator cannot distinguish. On the other hand, data distribution is approximated in VAE by using variational inference, where an encoder approximates the posterior distribution in the latent space and a decoder models the likelihood (\cite{Sabokrou2019AVID:Detection,Dehaene2020IterativeLocalization}).
Recent literature on unsupervised anomaly segmentation also includes non VAE and GAN based approaches. For instance, \citep{bergmann2020uninformed} exploits the teacher-student learning paradigm, highlighting anomalies on those outputs where the student networks and teacher model predictions differ. Additionally, feature-based methods \citep{shi2021unsupervised,bergmann2020uninformed}, which identify anomalies in the feature space can be also employed.

\subsection{Unsupervised anomaly segmentation in medical imaging}

In the context of medical images, most current literature resorts to VAEs, proposing several improvements to overcome specific limitations of simple VAEs \citep{Chen2018UnsupervisedAuto-encoders,NickPawlowski2018UnsupervisedAutoencoders,Chen2020UnsupervisedPrior,zimmerer2019context}. For example, to handle the lack of consistency in the learned latent representation on prior works, \cite{Chen2018UnsupervisedAuto-encoders} included a constraint that helps mapping an image containing abnormal anatomy close to its corresponding healthy image in the latent space. \cite{zimmerer2019context} presented a context-encoding VAE that combines reconstruction- with density-based anomaly scoring to capture the high-level structure present in the data. More recently, a probabilistic model that uses a network-based prior as the normative distribution on the latent-variable model was proposed in \citep{Chen2020UnsupervisedPrior}. In particular, this model penalized large deviations between the reconstructed and original input images, reducing false positives in pixel-wise predictions. Generative models have been also employed to tackle the unsupervised lesion segmentation task \citep{Baur2020SteGANomaly:MRI,nguyen2021unsupervised}. While SteGANomaly \citep{Baur2020SteGANomaly:MRI} integrated a CycleGAN-based style-transfer framework to map samples in the latent space much closer to the training distribution,  \cite{nguyen2021unsupervised} mask out random regions of the input data before they are fed to the GAN model. Note that a detailed survey on unsupervised anomaly localization in medical imaging can be found in \cite{Baur2021AutoencodersStudy}. However, despite the recent popularity of these methods, the results from the Medical Out-of-Distribution Analysis Challenge 2020 (\cite{MOODChallenge2020}) highlight their suboptimal performance on anomaly segmentation, which might impede their usability in clinical practice, as stressed by \cite{MeissenHistogram2}.

More recently, \cite{Venkataramanan2020AttentionImages} integrate attention maps derived from Grad-CAM (\cite{Selvaraju2020Grad-CAM:Localization}) during the training as supervisory signals. In particular, in addition to standard learning objectives, authors introduce an auxiliary loss that tries to maximize the attention maps on normal images by including an equality constraint with the form of a L$_1$ penalty over each individual pixel.

\subsection{Constrained segmentation}

Imposing global constraints on the output predictions of deep CNNs has gained attention recently, particularly in weakly supervised segmentation. These constraints can be embedded into the network outputs in the form of direct loss functions, which guide the network training when fully labeled images are not accessible. For example, a popular scenario is to enforce the softmax predictions to satisfy a prior knowledge on the size of the target region. \cite{Jia2017ConstrainedSegmentation} employed a L$_2$ penalty to impose equality constraints on the size of the target regions in the context of histopathology image segmentation. In \cite{Zhang2017CurriculumScenes}, authors leverage the target properties by enforcing the label distribution of predicted images to match an inferred label distribution of a given image, which is achieved with a KL-divergence term. Similarly, \cite{Zhou2019Prior-awareSegmentation} proposed a novel loss objective in the context of partially labeled images, which integrated an auxiliary term, based on a KL-divergence, to enforce that the average output size distributions of different organs approximates their empirical distributions, obtained from fully-labeled images. 

While the equality-constrained formulations proposed in these works are very interesting, they assume exact knowledge of the target size prior. In contrast, inequality constraints can relax this assumption, allowing much more flexibility. In \cite{Pathak2015ConstrainedSegmentation}, authors imposed inequality constraints on a latent distribution --which represents a “fake” ground truth-- instead of the network output, to avoid the computational complexity of directly using Lagrangian-dual optimization. Then, the network parameters are optimized to minimize the KL divergence between the network softmax probabilities and the latent distribution. Nevertheless, their formulation is limited to linear constraints. More recently, inequality constraints have been tackled by augmenting the learning objective with a penalty-based function, e.g., L$_2$ penalty, which can be imposed within a continuous optimization framework (\cite{Kervadec2019ConstrainedExtensions,Kervadec2019CurriculumSegmentation,Bateson2021ConstrainedSegmentation}), or in the discrete domain (\cite{Peng2020Discretely-constrainedSegmentation}). Despite these methods have demonstrated remarkable performance in weakly supervised segmentation, they require that prior knowledge, \textit{exact} or \textit{approximate}, is given. This contrasts with the proposed approach, which is trained on data without anomalies, and hence the size of the target is zero.


\section{Methodology}
\label{sec:methods}

\begin{figure*}[h!]
\begin{center}
\includegraphics[width=1\textwidth]{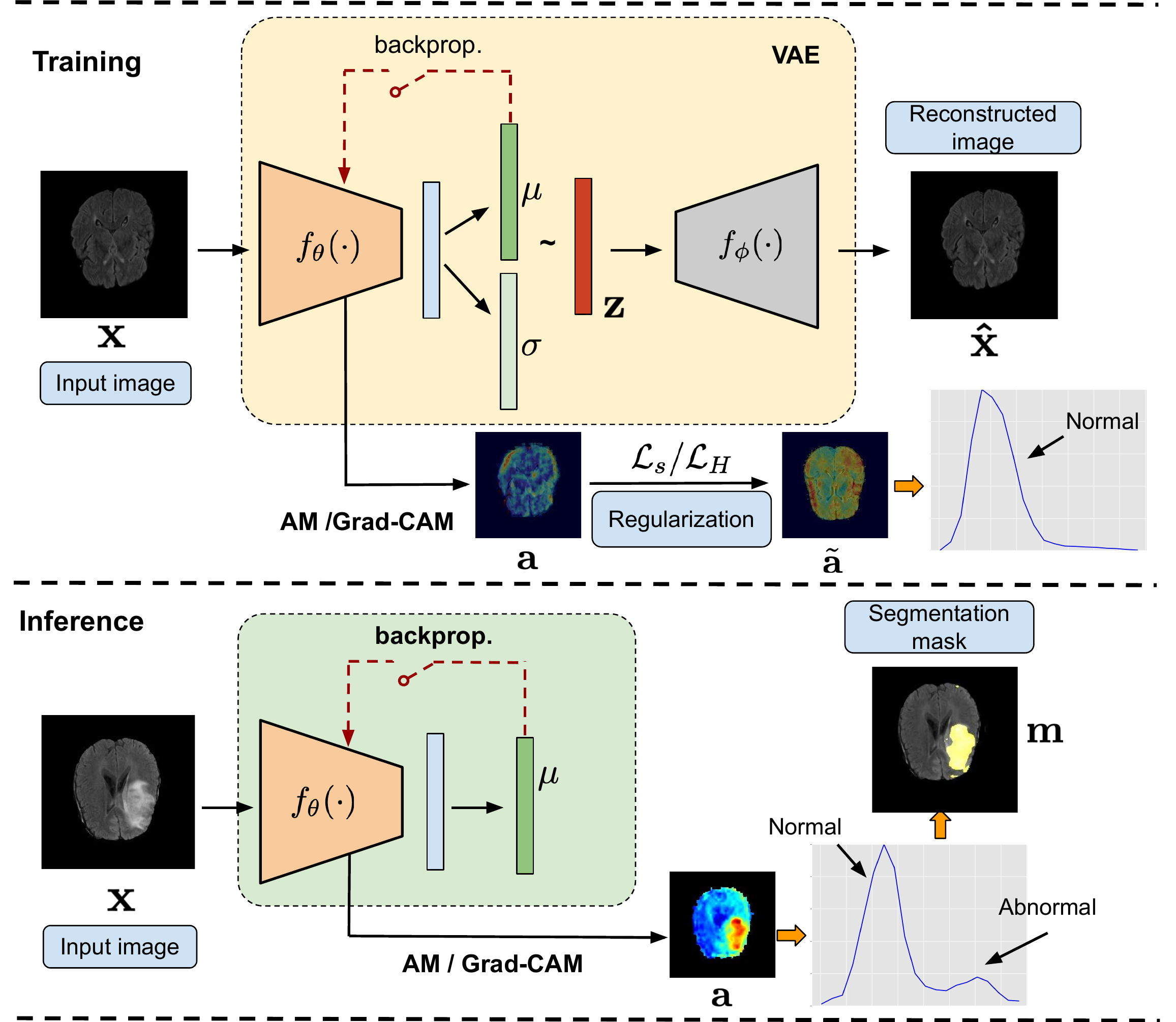}
\caption{\textbf{Method overview}. Following the standard literature, the VAE is optimized to maximize the evidence lower bound (ELBO), which satisfies Eq. \ref{eq:VAE_vanilla}. In addition, we include an attention constraint (in the form of a size-constrained loss $\mathcal{L}_s$ or entropy proxy $\mathcal{L}_H$) on the attention maps $\aaa$, to force the network to search in the whole image. At inference, the attention map is thresholded to obtain the final segmentation mask $\mm$.}
\label{fig:summary}
\end{center}
\vspace{-7mm}
\end{figure*}

An overview of our method is presented in Fig. \ref{fig:summary}. In what follows, we describe each component of our methodology.

\vspace{-2mm}
\paragraph{\textbf{Preliminaries}}Let us denote the set of unlabeled training images as $\mathcal{D} = \{\xx_n\}_{n=1}^N$, where $\xx_i \in \mathcal{X} \subset \real^{\Omega_i}$ represents the \textit{i}$^{th}$ image and $\Omega_i$ denotes the spatial image domain. This dataset contains only normal images, e.g., healthy images in the medical context, and has therefore no segmentation mask associated with each image. We now define an encoder, $f_{\ttt}(\cdot) : \mathcal{X} \rightarrow \mathcal{Z}$, parameterized by $\ttt$, which is optimized to project normal data points in $\mathcal{D}$ into a manifold represented by a lower dimensionality $d$, $\zz \in \mathcal{Z} \subset \real^{d}$. Furthermore, a decoder $f_{\ppp}(\cdot) : \mathcal{Z} \rightarrow \mathcal{X}$ parameterized by $\ppp$ aims at reconstructing an input image $\xx \in \mathcal{X}$ from $\zz \in \mathcal{Z}$, which results in $\hat \xx=f_{\ppp}(f_{\ttt}(\xx))$.

\subsection{Vanilla VAE}

A Variational Autoencoder (VAE) is an encoder-decoder style generative model, which is currently the dominant strategy for unsupervised anomaly location. Training a VAE consists in minimizing a two-term loss function, which is equivalent to maximize the evidence lower-bound (ELBO) (\cite{Kingma2014Auto-encodingBayes}):

\vspace{-2mm}

\begin{equation}
\mathcal{L}_{VAE} = \mathcal{L}_{R}(\xx, \hat{\xx}) + \beta\mathcal{L}_{KL}(q_{\ttt}(\zz|\xx) || p(\zz))
\label{eq:VAE_vanilla}
\end{equation}

\noindent where $\mathcal{L}_{R}$ is the reconstruction error term between the input and its reconstructed counterpart. The right-hand term is the Kullback-Leibler (KL) divergence (weighted by $\beta$) between the approximate posterior $q_{\ttt}(\zz|\xx)$ and the prior $p(\zz)$, which acts as a regularizer, penalizing approximations for $q_{\ttt}(\zz|\xx)$ that differ from the prior.

\subsection{Size regularizer via VAE attention}

Very recent literature (\cite{Liu2020TowardsAutoencoders,Venkataramanan2020AttentionImages}) has explored the use of attention maps for anomaly localization. In particular, attention maps $\aaa \in \real^{\Omega_i}$ are generated from the latent mean vector $\zz_{\mu}$, by using Grad-CAM (\cite{Selvaraju2020Grad-CAM:Localization}) via backpropagation to an encoder block output $f^s_{\ttt}(\xx)$, at a given network depth $s$. Thus, for a given input image $\xx^n$ its corresponding attention map is computed as follows:

\begin{equation}
\aaa^n= \sigma(\sum_{k}^{K} \alpha_{k} f^s_{\ttt}(\xx^n)_{k})
\label{eq:gradcams}
\end{equation}

\noindent where $K$ is the total number of filters of that encoder layer, $\sigma$ a sigmoid operation, and $\alpha_{k}$ are the generated gradients such that: $\alpha_{k}=\frac{1}{|\aaa^n|}\sum_{t \in \Omega_T}\frac{\partial \zz_{\mu}}{\partial \aaa^n_{k,t}}$, where $\Omega_T$ is the spatial features domain.

In \cite{Venkataramanan2020AttentionImages}, authors leveraged the Grad-CAMs based attention maps (Eq.\ref{eq:gradcams}) by enforcing them to cover the whole normal image. To achieve this, their loss function was augmented with an additional term, referred to as expansion loss, which takes the form of: $\mathcal{L}_s=\frac{1}{|\aaa|}\sum_{l \in \Omega_i}(1- \mathbf{a}^n_l)$. We can easily observe that this term resembles to multiple equality constraints, one at each pixel, forcing the class activation maps to be maximum at the whole image in a pixel-wise manner (i.e., it penalizes each single pixel individually). Contrary to this work, we integrate supervision on attention maps by enforcing inequality constraints on its global target size. Note that the use of the inequality constraints is motivated by the choice of the barrier function in the constrained problem, which is further detailed in Section \ref{ssec:logbarrier}. Hence, we aim at minimizing the following constrained optimization problem:

\begin{align}
\label{eq:constrained_eq}
\min_{\ttt,\ppp} \quad & \mathcal L_{VAE}(\ttt,\ppp) \qquad \text{s.t.} \quad f_c(\mathbf{a}^n) \leq 0, \quad n=1,...,N  
\end{align}

where $f_c(\mathbf{a}^j)=(1- \frac{1}{|\Omega_i|}\sum_{l \in \Omega_i} \mathbf{a}^n_l)$ is the constraint over the attention map from the $j$-\textit{th} image, which enforces the generated attention map to cover the whole image. It is well-known in optimization that a penalty does not act as a barrier near the boundary of the feasible set \citep{boyd2004convex}. In other words, a constraint that is satisfied results in a null penalty and gradient. Therefore, at a given gradient update, there is nothing that prevents a satisfied constraint from being violated, causing oscillations between competing constraints and ultimately resulting in a potential unstable training. This is further exacerbated in the case of many multiple constraints (i.e., \cite{Venkataramanan2020AttentionImages}), motivating the use of a single global constraint to achieve a maximum coverage of class-activation maps over the whole image in our scenario. From Eq. \ref{eq:constrained_eq} we can derive an approximate unconstrained optimization problem by employing a penalty-based method, which takes the hard constraint and moves it into the loss function as a penalty term ($\mathcal{P}(\cdot)$): $\min_{\ttt,\ppp} \mathcal{L}_{VAE}(\ttt,\ppp) + \lambda \mathcal{P}(f_c(\mathbf{a}))$. Thus, each time that the constraint $f_c(\mathbf{a}^n) \leq 0$ is violated, the penalty term $\mathcal{P}(f_c(\mathbf{a}^n))$ increases.

\subsection{Extended log-barrier as an alternative to penalty-based functions}
\label{ssec:logbarrier}

Despite having demonstrated a good performance in several applications (\cite{Kervadec2019Constrained-CNNSegmentation,Pathak2015ConstrainedSegmentation,He2017LearningTightening,Jia2017ConstrainedSegmentation}) penalty-based methods have several drawbacks. First, these unconstrained minimization problems have increasingly unfavorable structure due to ill-conditioning (\cite{FiaccoAnthonyVandMcCormick1990NonlinearTechniques,Luenberger1973IntroductionProgramming}), which typically results in an exceedingly slow convergence. Second, finding the optimal penalty weight is not trivial. In addition, we advocate for the use of the log-barrier extension versus penalties due to the strictly positive gradient of the latter becomes higher when a satisfied constraint approaches violation during optimization, pushing it back towards the feasible set (See Figure 1 in \cite{Kervadec2019ConstrainedExtensions}). As explained in the previous section, this contrasts with penalties, as they deliver null gradients if a given constraint is satisfied. To address these limitations, we replace the penalty-based functions by the approximation of log-barrier\footnote{Note that this function is convex, continuous and twice-differentiable.} presented in \cite{Kervadec2019ConstrainedExtensions}. We would like to stress that barrier methods require the interior of the feasible sets to be non-empty and they are used, therefore, in constrained optimization problems with inequality constraints, such as the one defined in Eq. \ref{eq:constrained_eq} (note that there is no interior for equality constraints). Thus, we can formally define the approximation of log-barrier as:

\begin{equation}
\label{eq:log_barrier_extension}
\widetilde{\psi}_{t}(z) =
\begin{cases}
-\frac{1}{t} \log (-z) & \text{if } z \leq -\frac{1}{t^2} \\
tz - \frac{1}{t} \log (\frac{1}{t^2}) + \frac{1}{t} & \text{otherwise} ,
\end{cases}
\end{equation}

where $t$ \textit{controls} the barrier during training, and $z$ is the constraint $f_c(\mathbf{a}^n)$. Thus, by taking into account the approximation in \ref{eq:log_barrier_extension}, we can solve the following unconstrained problem by using standard Gradient Descent:

\vspace{-2mm}
\begin{equation}
\label{eq:criterion}
\min_{\ttt,\ppp} \quad  \underbrace{\mathcal L_{VAE}(\ttt,\ppp)}_{\text{Standard VAE loss}} + \lambda_{s} \underbrace{\sum_{n=1}^N\widetilde{\psi}_{t}(1- \frac{1}{|\Omega_i|}\sum_{l \in \Omega_i} \mathbf{a}^n_l)}_{\mathcal{L}_s:~\text{Size regularizer}}
\end{equation}

In this scenario, for a given $t$, the optimizer will try to find a solution with a good compromise between minimizing the loss of the VAE and satisfying the constraint $f_c(\mathbf{a}^n)$. In the following, we refer to this formulation of gradient-CAM constraint as GradCAMCons setting.

\subsection{On the role of gradients in VAEs}

Even though there exist a few initial attempts to integrate attention maps on the task of unsupervised anomaly detection, how gradient-based attention behave on anomalous patterns remains unclear. For instance, \cite{Liu2020TowardsAutoencoders} argue that anomalies produce larger gradients in the learned latent representation, which results in higher activated attention maps. On the other hand, \cite{Venkataramanan2020AttentionImages} states that the VAE only focus on normal patterns (with which it has been trained), thus anomalous regions produce smaller absolute value gradients. These inconsistencies in the literature have motivated us to analyze the underlying role of the gradients in the context of brain images analysis. Thus, we performed several experiments to analyze the behaviour of grad-CAMs in anomaly localization compared to non-weighted activation maps (AMs), which are computed as:

\begin{equation}
\aaa^n= \frac{1}{K}\sum_{k}^{K} f^s_{\ttt}(\xx^n)_{k}
\label{eq:am}
\end{equation}

\begin{figure}[h!]
    \begin{center}

          \includegraphics[width=.8 \linewidth]{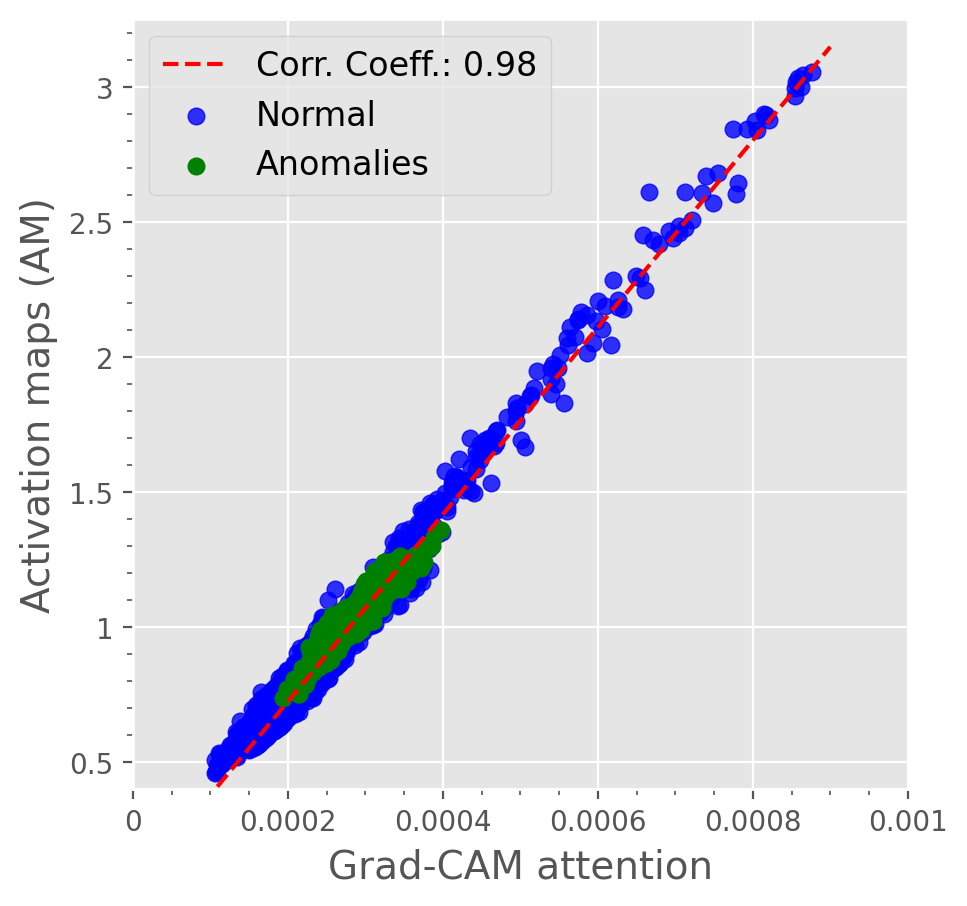}
         
        \caption{Relation between the activation values and gradient-weighted attention maps in an unconstrained VAE. These results demonstrate that the values obtained by Grad-CAM based attention are highly correlated (correlation coefficient = 0.98) to those obtained by the attention maps, suggesting that the gradient basically contributes as a scaling factor on the attention maps.}
        \label{fig:am_vs_gradcam}
    \end{center}
    \vspace{-2 em}
\end{figure}

In particular, we could not find any benefit on gradients weighting other than serving as a scaling factor for attention maps to fall on non-saturated range of values of typically used activation functions, such as the sigmoid operation in Eq. \ref{eq:gradcams} (see Figure \ref{fig:am_vs_gradcam}, where we show that the values obtained by both types of attention are highly correlated). Furthermore, we found that the reconstructed images derived from the gradient-based attention contained more errors compared to those reconstructed with attention on the activation maps (Eq \ref{eq:am}). We refer the reader to Section 1 of Supplemental Material for the detailed results concerning the role of the gradients.

\subsection{Entropy maximization as a proxy for the constraint}

Based on our previous findings, we advocate that the use of non-weighted activation maps (AMs) should be preferred over their gradient-based counterpart. Nevertheless, this solution has a main limitation that hinders the use of size constraints. As the activation maps are not normalized, the arbitrary activation value to impose the constraint loses the sense of \textit{size} or \textit{proportion}. The activation values produced by neural networks can vary in each application, as well as with the architecture used, which makes it difficult to establish generalizable restrictions on their value. For this reason, we propose to use attention maps derived from normalizing the activation maps over all the pixels of the image, via a softmax activation, similarly to \cite{Ilse2018Attention-basedLearning}, such that: $p^{n} = \tau_{\Omega_B}(\mathbf{a}^n)$\footnote{Note that $\tau$ is the softmax activation on the brain tissue instances, $\Omega_B$.}. Since these attention maps are normalized across pixels and not over classes, the use of global constraints is meaningless, as the sum over all the pixels post-softmax will be equal to 1.0. Nevertheless, we still aim at regularizing the attention distribution $p^{n}$ to focus on all patterns in the image \textit{homogeneously}. To this end, we propose to minimize the KL distance $D_{KL}(p||q)=H(p,q)-H(p)$ between the attention distribution $p$, and a constant distribution $q$, where $H(p,q)$ represents the cross-entropy between both distributions, and $H(p)=H(p,p)$ is the Shannon entropy of the intensity distribution such that $H(p)=-\frac{1}{I}\sum_{i}p_{i} \cdot log(p_{i})$. In the scenario where we want $p$ to match a constant distribution, it is straightforward to see that minimizing the KL distance is equivalent to maximizing the entropy $H(p)$:

\vspace{-2mm}
\begin{equation}
\label{eq:kl}
D_{KL}(p||q)=H(p,q)-H(p)=^{c}-H(p)
\end{equation}

\noindent where $=^{c}$ indicates equality up to an additive constant.

Thus, the proposed constrained optimization problem integrating an entropy maximization term, referred to as $\mathcal{L}_H$, offers a softer attention constraint compared to the solution in Eq. \ref{eq:criterion}. Furthermore, this formulation allows the VAE to keep the most suitable activation values, while requiring less hyper-parameters to be optimized. Analogously to Eq. \ref{eq:criterion}, we solve the constrained optimization problem with $\mathcal{L}_H$ by using standard Gradient Descent:

\begin{equation}
\label{eq:criterion_Lh}
\min_{\ttt,\ppp} \quad  \underbrace{\mathcal L_{VAE}(\ttt,\ppp)}_{\text{Standard VAE loss}} - \lambda_{H} \underbrace{\frac{1}{N}\sum_{n=1}^N H(\tau_{\Omega_B}(\mathbf{a}^n))}_{\mathcal{L}_H:~\text{Entropy regularizer}}
\end{equation}

Hereafter, we will refer to this formulation as AMCons. 

\subsection{\textbf{Inference}} 

During inference, we use the generated attention as an anomaly saliency map. For the Grad-CAMs based settings we replaced the sigmoid operation by a minimum-maximum normalization in order to avoid saturation caused by large activations. During the experimental stage, we found that anomalies produce larger activation on attention maps than the constrained normal samples, in line to prior literature (\cite{Liu2020TowardsAutoencoders}). Then, the map is thresholded to create an anomaly mask of the image.


\section{Experimental setting}
\label{sec:exp}

\subsection{Datasets}

The experiments described in this work are carried out in the context of brain lesions localization. Concretely, we use two relevant neuroimaging challenges: tumour segmentation in MRI volumes and intracranial hemorrhage (ICH) segmentation in CT scans.

\paragraph{\textbf{Brain tumor segmentation}} For this task, we used the popular BraTS 2019 dataset (\cite{Menze2015TheBRATSJ, Bakas2017AdvancingFeaturesJ, Bakas2018IdentifyingChallengeJ}), which contains $335$ multi-institutional multi-modal MR scans with their corresponding Glioma segmentation masks. Following \cite{Baur2019DeepImages}, from every patient, $10$ consecutive axial slices of FLAIR modality of resolution $224\times224$ pixels were extracted around the center to get a pseudo MRI volume. Then, the dataset is split into training, validation and testing groups, with $271$, $32$ and $32$ patients, respectively. Following the standard literature, during training only the slices without lesions are used as normal samples. For validation and testing, scans with less than $0.01\%$ of tumour are discarded, following the standard practices in the literature. 

\paragraph{\textbf{Intracranial hemorrhage segmentation}} We use the Physionet-ICH dataset (\cite{Hssayeni2020ComputedSegmentation_blue, Hssayeni2020IntracranialModel_blue, Goldberger2000PhysioBankSignals_blue}) to localize intracranial hemorrhage lesions. The dataset is composed of $82$ non-contrast CT scans of subjects with traumatic brain injury. From those, $36$ cases are diagnosed with intracranial hemorrhage of different types: Intraventricular, Intraparenchymal, Subarachnoid, Epidural and Subdural. ICH Lesions were slice-wise delineated by two expert radiologists. In our work, we join the different ICH types into one single label for binary lesion segmentation. CT scans are skull-stripped, intensity-normalized, and co-registered into a reference scan. Similar to the BraTS dataset, $10$ consecutive axial slices of resolution $224\times224$ pixels around the center were extracted to get CT pseudo volumes. The dataset is divided into training, validation and testing splits. The first one contains only non-ICH cases (n=46), while cases with labeled lesions were used for validation (n=6) and testing (n=30). Although the main core of ablation experiments in this work are described on the BraTS dataset, we use the Physionet-ICH dataset to demonstrate the generalization capabilities of our proposed method on different brain lesions and imaging modalities.

\subsection{Evaluation Metrics}

We resort to standard metrics for unsupervised brain lesion segmentation, as in \cite{Baur2021AutoencodersStudy}. Concretely, we compute the dataset-level area under precision-recall curve (AUPRC) at pixel level, as well the area under receptive-operative curve (AUROC). From the former, we obtain the operative point (OP) as threshold to generate the final segmentation masks. Then, we compute the best dataset-level S\o{}rensen-Dice score ($\lceil$DICE$\rceil$) and intersection-over-union ($\lceil$IoU$\rceil$) over these segmentation masks. Finally, we compute the average S\o{}rensen-Dice score (DICE) over single scans. For each experiment, the metrics reported are the average of three consecutive repetitions of the training, to account for the variability of the stochastic factors involved in the process.

\subsection{Implementation Details}

The VAE architecture used in this work is based on the recently proposed framework in \cite{Venkataramanan2020AttentionImages}. Concretely, the convolution layers of ResNet-18 (\cite{He2016DeepRecognition}) are used as the encoder, followed by a dense latent space $\zz\in \real^{32}$. For image generation, a residual decoder is used, which is symmetrical to the encoder. It is noteworthy to mention that, even though several methods have resorted to a spatial latent space (\cite{Baur2019DeepImages,Venkataramanan2020AttentionImages}), we observed that a dense latent space provided better results, which aligns to the recent benchmark in \cite{Baur2021AutoencodersStudy}. To train the GradCAMCons formulation in eq. \ref{eq:criterion} we first trained the VAE during $50$ epochs without any expansion to stabilize the convergence using $\beta = 1$. Then, the proposed regularizer was integrated (equation \ref{eq:criterion}) with $t=10$ and $\lambda_{s}=10^3$ applied to the Grad-CAMs obtained from the first convolutional block of the encoder during $250$ epochs. We use a batch size of $8$ images, and a learning rate of $1e{-5}$ with ADAM as optimizer. The reconstruction loss, $\mathcal{L}_{R}$, in eq. (\ref{eq:VAE_vanilla}) is the binary cross-entropy. Similarly, the AMCons formulation in eq. \ref{eq:criterion_Lh} was trained by using $\beta = 10$ and $\lambda_{H} = 0.1$, using a learning rate of $1e{-4}$. Ablation experiments to motivate the choice of values used are presented in Section \ref{sec:ablation} and Section 3 of supplemental materials. The code and trained models are publicly available  on (\url{https://github.com/jusiro/constrained_anomaly_segmentation/}).  

\vspace{-3mm}

\subsection{Baselines}

In order to compare our approach to state-of-the-art methods, we implemented prior works and validated them on the dataset used, under the same conditions. First, we use residual-based methods to match the recently benchmark on unsupervised lesion localization in \cite{Baur2021AutoencodersStudy}. Then, we implement up-to-date methods based on contrast adjustment on the input image via histogram equalization. We also include recently proposed methods that integrate CAMs to locate anomalies. For both strategies, the AE/VAE architecture was the same as the one used in the proposed method. \textbf{\textit{Residual methods}}, given an anomalous sample, aim to use the AE/VAE to reconstruct its normal counterpart. Then, they obtain an anomaly localization map using the residual between both images such that $\mm = \lvert \xx -\hat{\xx} \rvert$, where $\lvert \cdot \rvert$ indicates the absolute value. On the AE/VAE scenario, we include methods which propose modifications over vanilla versions, including context data augmentation in Context AE \cite{zimmerer2019context}, Bayesian AEs (\cite{NickPawlowski2018UnsupervisedAutoencoders}), Restoration VAEs (\cite{Chen2020UnsupervisedPrior}), an adversarial-based VAEs, AnoVAEGAN (\cite{Baur2019DeepImages}) and a recent GAN-based approach, F-anoGAN (\cite{Schlegl2019F-AnoGAN:Networks}). For methods including adversarial learning, DC-GAN \cite{Radford2016UnsupervisedNetworks} is used as discriminator. During inference, residual maps are masked using a slight-eroded brain mask, to avoid noisy reconstructions along the brain borderline. \textbf{\textit{Equalization-based methods}}: very recent methods have highlighted the limits of residual-based approaches to properly discern brain lesions \cite{MeissenHistogram, MeissenHistogram2}. In contrast, they propose to apply an equalization of the histogram of the input image, and to set a threshold on the preprocessed image, considering that brain lesions often show hyperintense patterns in different modalities. Concretely, we include the method proposed in \cite{MeissenHistogram}, which we refer to as HistEq. \textbf{\textit{CAMs-based}}: we use Grad-CAM VAE (\cite{Liu2020TowardsAutoencoders}), which obtains regular Grad-CAMs on the encoder from the latent space $\zz_{\mu}$ of a trained vanilla VAE. Concretely, we include a disentanglement variant of CAMs proposed in this work, which computes the combination of individually-calculated CAMs from each dimension in $\zz_{\mu}$, referred to as Grad-CAM$_{D}$ VAE. We also use the recent method in \cite{Venkataramanan2020AttentionImages} (CAVGA), which applies a L1 penalty on the generated CAM to maximize the attention. In contrast to our model and \cite{Liu2020TowardsAutoencoders}, the anomaly mask in \cite{Venkataramanan2020AttentionImages} is generated by focusing on the regions not activated on the saliency map such that $\aaa = 1 - CAM$, hypothesizing that the network has learnt to focus only on normal regions. Then, $\aaa$ is thresholded with 0.5 to obtain the final anomaly mask $\mm \in \mathbb{R}^{\Omega_i}$. For both methods, the network layer to obtain the Grad-CAMs is the same as in our method. 

\section{Results}
\label{sec:results}


\subsection{Comparison to the literature.}
\label{results_literature}

The quantitative results obtained by the proposed model and baselines on the test cohort are presented in Table \ref{tab:test_results}. Results from residual-based baselines range between {[}$0.056$-$0.511${]}(AUPRC) and {[}$0.188$-$0.525${]} (DICE), which are in line with previous literature \cite{Baur2021AutoencodersStudy}. We can observe that the proposed formulations outperform these approaches by a large margin. Concretely, the AMCons method provides a substantial increase of $\sim$34\% and $\sim$26\% in terms of AUPRC and DICE, respectively, compared to the best model, i.e., F-anoGAN. Furthermore, the model integrating the $\mathcal{L}_H$ term significantly outperforms our previous method in \cite{Silva-Rodriguez2021LookingSegmentation}. This supports our hypothesis that using non-weighted attention maps with a maximization entropy term as constraint is indeed a better solution for the unsupervised lesion segmentation task. Finally, in comparison with the very recently proposed method of histogram equalization, HistEq, our proposed formulation brings improvements of nearly $\sim$10\% in the main figures of merit.

\begin{table*}[h!]
\resizebox{\textwidth}{!}{%
\centering
\small
\begin{tabular}{|l|c|c|c|c|c|}
\hline
\multicolumn{1}{|c|}{\textbf{Method}} &  \multicolumn{1}{c|}{\textbf{AUROC}} & \multicolumn{1}{c|}{\textbf{AUPRC}} & \multicolumn{1}{c|}{\textbf{$\lceil$DICE$\rceil$}} & \multicolumn{1}{c|}{\textbf{$\lceil$IoU$\rceil$}} & \multicolumn{1}{c|}{\textbf{DICE ($\mu\pm\sigma$)}} \\ \hline\hline
CAVGA (\cite{Venkataramanan2020AttentionImages})   &$0.726$($0.001$) & $0.056$($0.005$) & $0.188$($0.001$) & $0.104$($0.002$) & $0.182$($0.004$)$\pm0.096$($0.002$) \\ \hline
Bayesian VAE (\cite{NickPawlowski2018UnsupervisedAutoencoders}) &  $0.922$($0.002$) & $0.193$($0.005$) & $0.342$($0.005$) & $0.206$($0.005$) & $0.329$($0.005$)$\pm0.115$($0.005$) \\ \hline
AnoVAEGAN (\cite{Baur2019DeepImages}) &  $0.925$($0.020$) & $0.232$($0.052$) & $0.359$($0.074$) & $0.221$($0.053$) & $0.349$($0.071$)$\pm0.115$($0.015$) \\ \hline
Bayesian AE (\cite{NickPawlowski2018UnsupervisedAutoencoders}) &  $0.940$($0.002$) & $0.279$($0.009$) & $0.389$($0.012$) & $0.242$($0.009$) & $0.375$($0.010$)$\pm0.130$($0.011$) \\ \hline
AE &  $0.937$($0.002$) & $0.261$($0.011$) & $0.397$($0.011$) & $0.248$($0.008$) & $0.386$($0.010$)$\pm0.125$($0.004$)\\ \hline
Grad-CAM$_{D}$ VAE (\cite{Liu2020TowardsAutoencoders})  & $0.941$($0.003$)  & $0.312$($0.010$) & $0.400$($0.009$) & $0.250$($0.012$) & $0.361$($0.014$)$\pm0.164$($0.005$) \\ \hline
Restoration VAE (\cite{Chen2020UnsupervisedPrior})  & $0.934$($0.028$) & $0.352$($0.111$) & $0.403$($0.099$) & $0.252$($0.069$) & $0.345$($0.075$)$\pm0.186$($0.044$) \\ \hline
Context VAE (\cite{zimmerer2019context}) & $0.939$($0.004$) & $0.271$($0.017$) & $0.406$($0.020$) & $0.255$($0.016$) & $0.394$($0.017$)$\pm0.126$($0.007$) \\ \hline
Context AE (\cite{zimmerer2019context}) & $0.940$($0.003$) & $0.278$($0.012$) & $0.411$($0.014$) & $0.259$($0.011$) & $0.399$($0.013$)$\pm0.126$($0.005$)\\ \hline
VAE (\cite{Baur2019DeepImages,Zimmerer2020Abstract:Auto-encoders}) & $0.940$($0.002$) & $0.273$($0.010$) & $0.411$($0.012$) & $0.259$($0.009$) & $0.399$($0.010$)$\pm0.127$($0.004$) \\ \hline
F-anoGAN (\cite{Schlegl2019F-AnoGAN:Networks}) & $0.946$($0.026$) & $0.511$($0.190$) & $0.525$($0.147$) & $0.369$($0.131$) & $0.494$($0.138$)$\pm0.151$($0.038$)  \\ \hline
\rowcolor{gray!5} GradCAMCons w. $\mathcal{L}_S$ (L2 penalty) & $0.969$($0.015$) & $0.567$($0.138$) & $0.620$($0.085$) & $0.455$($0.086$) & $0.586$($0.079$)$\pm0.184$($0.028$) \\ \hline
HistEq (\cite{MeissenHistogram}) & $0.972$($0.000$) & $0.725$($0.000$) & $0.705$($0.000$) & $0.545$($0.000$) & $0.653$($0.000$)$\pm0.233$($0.000$) \\ \hline
\rowcolor{gray!5} GradCAMCons w. $\mathcal{L}_S$ (Log Barrier) & $0.982$($0.001$) & $0.746$($0.034$) & $0.698$($0.034$) & $0.537$($0.041$) & $0.677$($0.021$)$\pm0.215$($0.019$) \\ \hline
\rowcolor{gray!5}\textbf{AMCons w. $\mathcal{L}_H$} & $\mathbf{0.988}$($0.000$) & $\mathbf{0.850}$($0.011$) & $\mathbf{0.786}$($0.009$) & $\mathbf{0.648}$($0.013$) & $\mathbf{0.741}$($0.009$)$\mathbf{\pm0.153}$($0.001$) \\ \hline

\end{tabular}
}
\caption{Comparison to prior literature on BraTS dataset.  Results derived from the proposed methods in gray. Best results in bold. The values in parentheses indicate the standard deviation over the three training repetitions.}
\label{tab:test_results}
\vspace{-5mm}
\end{table*}

\subsection{Ablation experiments}
\label{sec:ablation}

The following ablation studies aim at demonstrating, in an empirical way, the motivation of employing the proposed models. First, we provide quantitative evidences about the better performance of using global constraints (model in Eq. \ref{eq:criterion}) over pixel-level constraints (i.e., \cite{Venkataramanan2020AttentionImages}). Second, we show that resorting to the extended log-barrier function is a better alternative than standard L2 penalty functions. Then, we perform an in-depth analysis of the optimal hyperparameters values for the entropy-guided model (Eq. \ref{eq:criterion_Lh}), as well as other important design choices.
\vspace{-1. em}
\paragraph{\textbf{Image vs. pixel-level constraint}} The following experiment demonstrates the benefits of imposing the constraint on the whole image rather than in a pixel-wise manner, such as in \cite{Venkataramanan2020AttentionImages}. In particular, we compare the two strategies when the constraint is enforced via a L2-penalty function, whose results are presented in Table \ref{fig:ablation_Ls_constraint}. In particular, we can easily see that imposing the constraint at image-level consistently outperforms pixel-level constraints. These results support our hypothesis that global constraints, such as the proposed formulation in Eq. \ref{eq:criterion}, should be preferred over multiple pixel-wise constraints, similar to \cite{Venkataramanan2020AttentionImages}.

\begin{table}[h!]
\centering
\footnotesize
\begin{tabular}{|l|c|c|c|}
\hline
\multicolumn{1}{|c|}{} &
  \begin{tabular}[c]{@{}c@{}}L2\\ (pixel-level)\end{tabular} &
  \begin{tabular}[c]{@{}c@{}}L2\\ (image-level)\end{tabular} &
  \begin{tabular}[c]{@{}c@{}}Log-Barrier\\ (image-level)\end{tabular} \\ \hline
AUPRC &
  0.489(0.098) &
  0.550(0.160) &
  0.728(0.034) \\ \hline
\end{tabular}
\caption{Quantitative comparison, in terms of AUPRC, between enforcing the constraint at pixel-level (i.e., \cite{Venkataramanan2020AttentionImages}) or at image-level (i.e., proposed approach), and for the impact of the type of regularization.}
 \label{fig:ablation_Ls_constraint}
\end{table}

\vspace{-1. em}


\paragraph{\textbf{Extended log-barrier vs. penalty-based functions}}To motivate the choice of employing the extended log-barrier over standard penalty-based functions in the constrained optimization problem in Eq. (\ref{eq:constrained_eq}), we compare them in Table \ref{fig:ablation_Ls_constraint}. It can be observed that imposing the constraint with the extended log-barrier consistently outperforms the L2-penalty, with substantial performance gains.

\begin{figure}[h!]
    \begin{center}

          \subfloat[\label{fig:ablation_H_1}]{\includegraphics[width=0.8\linewidth]{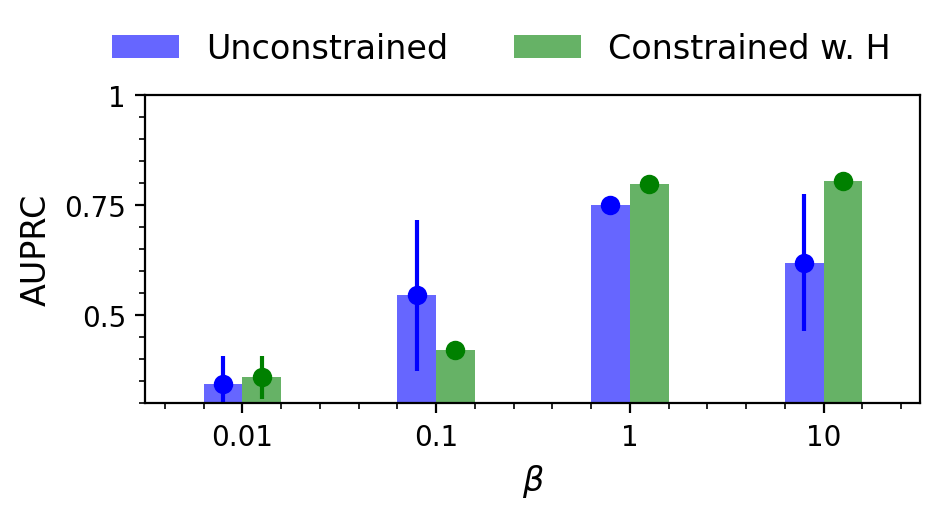}}
          \vspace{-1em}
          \subfloat[\label{fig:ablation_H_2}]{\includegraphics[width=0.8\linewidth]{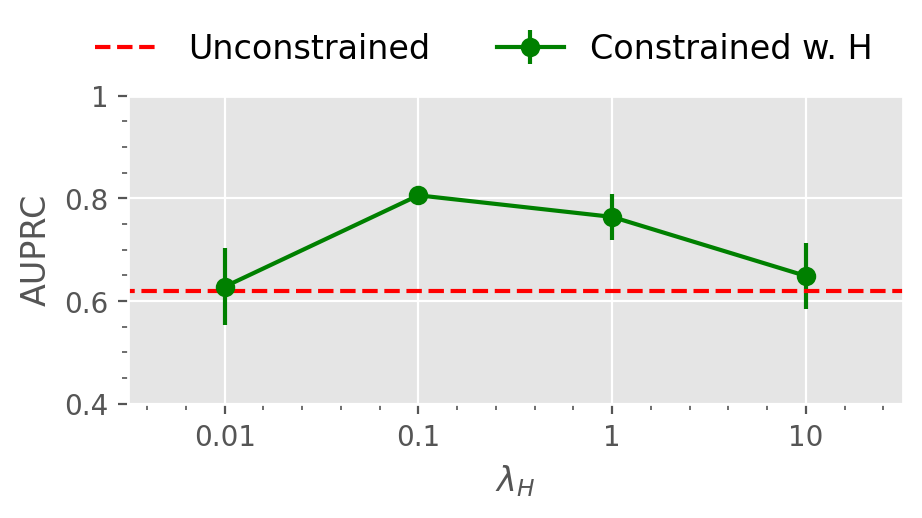}}

        \caption{Ablation study on the AMCons setting. Concretely, the role of the KL regularization ($\beta$) in the VAE and the entropy constraint on attention maps ($\lambda_{H}$) from our formulation is studied. (a) Entropy constraint effect and dependency on $\beta$. (b) Ablation study on $\lambda_{H}$.}
        \label{fig:ablation_H}
    \end{center}
    \vspace{-2.15 em}
\end{figure}

\paragraph{\textbf{On the impact of entropy-guided constraints}} We now perform an in-depth analysis of the effect of integrating the entropy-guided constraint in Eq. \ref{eq:criterion_Lh} for anomaly localization, as well as an extensive validation of the values of the balancing terms $\beta$ and $\lambda_{H}$. First, we study the impact of $\mathcal{L}_{H}$ across different $\beta$ values (i.e. $\beta=\{0.01, 0.1, 1, 10\}$), by fixing its balancing term $\lambda_{H}$ to 0.1, a value that empirically showed good stability. These results, which are reported in Figure \ref{fig:ablation_H_1}, show that the VAE with and without entropy constraint presents different optimal values for $\beta$. Nevertheless, the best results are obtained when the contribution of the regularization term is large (i.e. $\beta \geq 1$), and the entropy-based regularization over the activation maps included (i.e., green bars). Furthermore, this configuration is shown to be more stable once a large $\beta$ weight is set, particularly for the constrained formulation. Then, based on the best configuration ($\beta=10$), we study how different $\lambda_{H}$ weights $\{0.01, 0.1, 1, 10\}$ impact the model performance. These results (Figure \ref{fig:ablation_H_2}) show that incorporating the entropy regularization always contributes to performance gains, with an optimum weight value of $\lambda_{H}=0.1$.

\begin{figure}[h!]
    \begin{center}
    \includegraphics[width=1\linewidth]{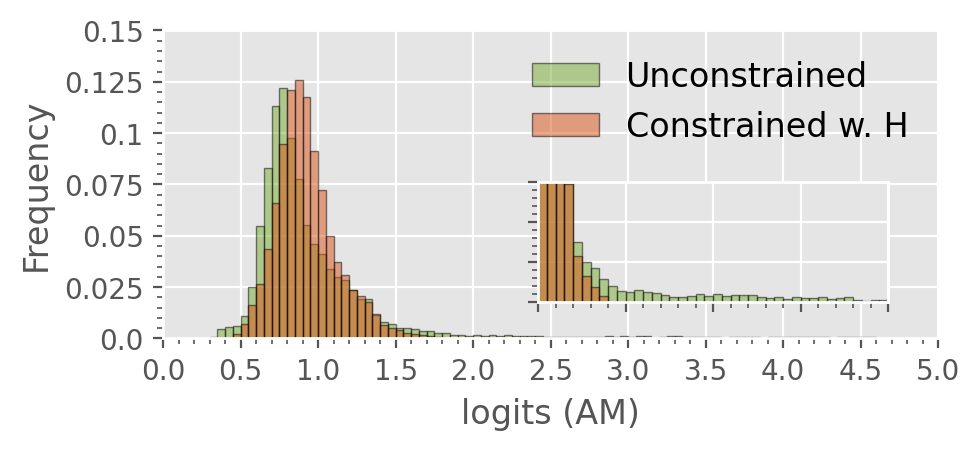}
    \end{center}
    \vspace{-1 em}
    \caption{Influence of the entropy constrained term on the attention maps for AMCons on normal images.}
     \label{fig:effectH_normal}
\end{figure}

\begin{figure}[h!]
    \begin{center}
    \includegraphics[width=.95\linewidth]{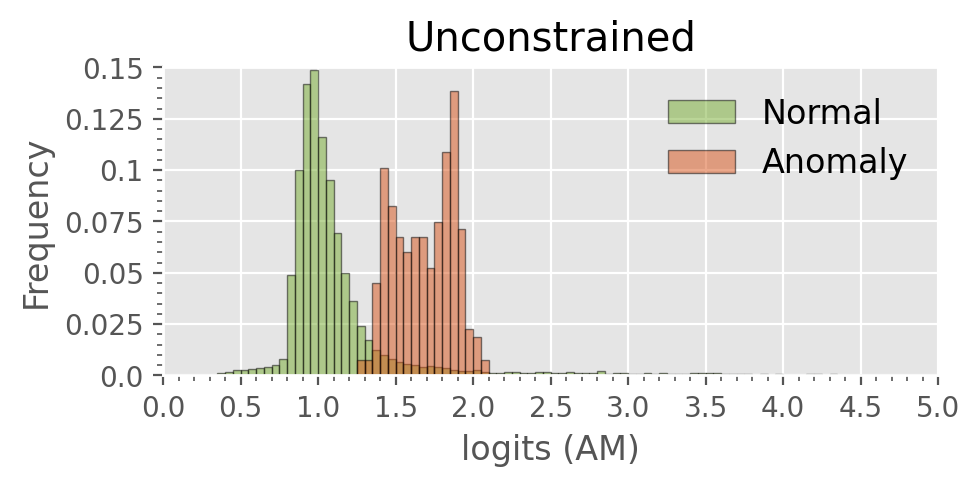}
    \includegraphics[width=.95\linewidth]{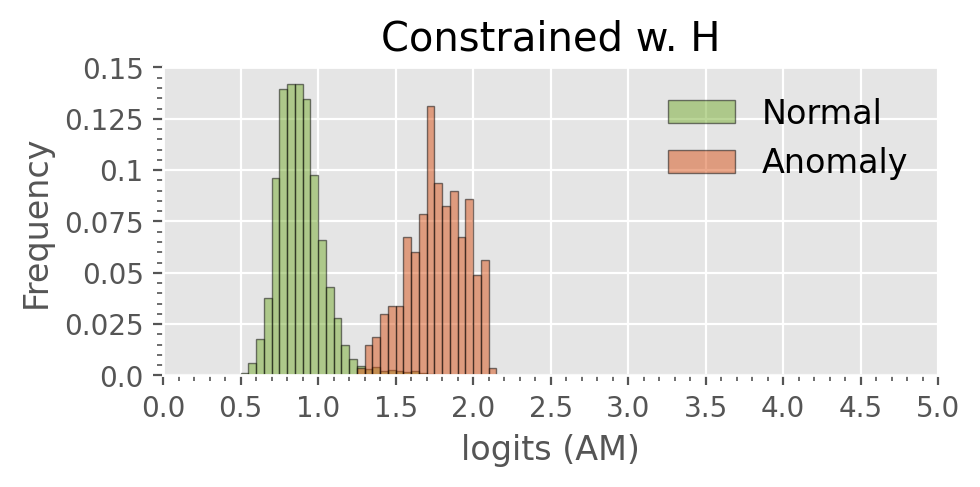}
    \end{center}
    \vspace{-2 em}
    \caption{Influence of the entropy constrained term on the attention maps for AMCons on images with anomalies.}
    \label{fig:effectH_abnormal}
\end{figure}

In the next experiment, we show how adding the $\mathcal{L}_{H}$ term in our formulation impacts the activation maps (AM). Concretely, we first show in Figure \ref{fig:effectH_normal} the AM distribution for a normal sample for both the constrained and unconstrained configurations. It can be observed that, in our constrained formulation, the distribution of activation values is more homogeneous (in orange), unlike the more spread values found in its unconstrained counterpart (in green). Furthermore, we show its impact on unseen, anomalous samples, where the benefits of our model are better highlighted. In particular, we represent the AM distribution for normal and anomalous pixels on the unconstrained formulation (i.e. $\lambda_{H} = 0$) in Figure \ref{fig:effectH_abnormal} (\textit{top}), and the effect of integrating the $L_{H}$ term (Figure \ref{fig:effectH_abnormal}, \textit{bottom}). Similarly to the normal samples, the distribution of normal pixels produced by the unconstrained setting spreads over a larger range, resulting in a higher overlapping with the distribution of anomalous pixels. Note that, in addition to the overlapping regions, there exist values of normal pixels which overpass anomalous values. In contrast, the more compact distribution provided by the proposed formulation favors a smaller overlap between normal and anomalous pixel intensity distributions. This results in an easier identification of normal \textit{versus} anomalous pixels.

\begin{figure}[h!]
    \begin{center}
    \includegraphics[width=1\linewidth]{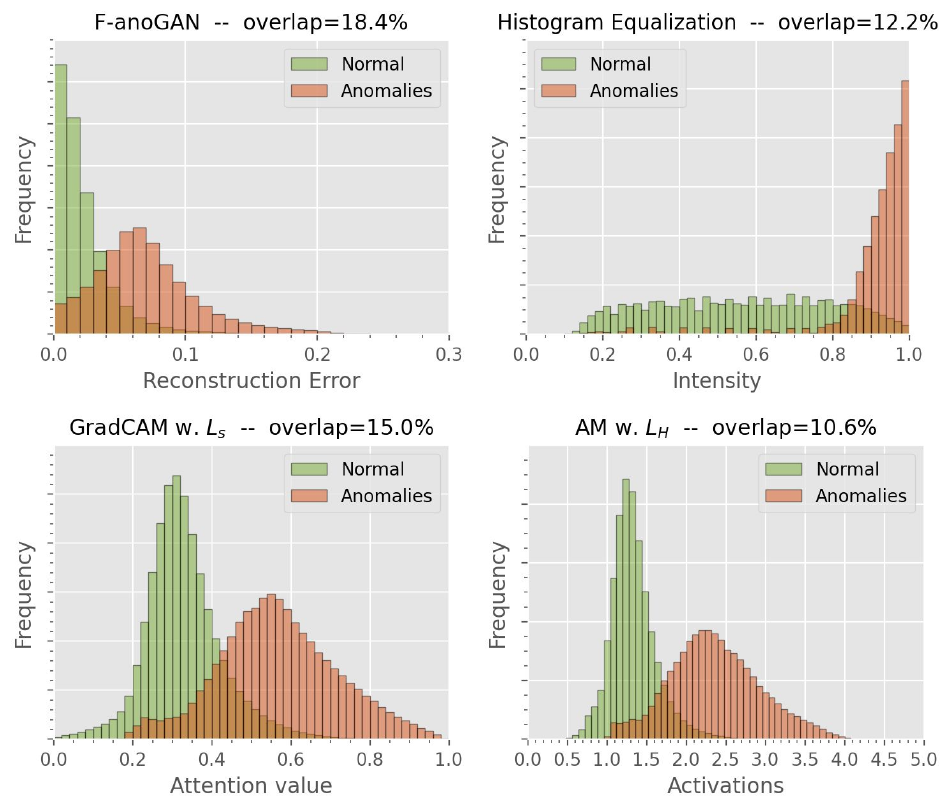}
    \end{center}
    \vspace{-1 em}
    \caption{Histogram analysis on the overlap of normal and anomalous samples for the different proposed methods and baselines (on the whole BRATS dataset).}
     \label{fig:overlap}
\end{figure}

\vspace{1 em}
In the following, we explore how the entropy constraint favors the smallest overlap between normal and anomalous distribution on the objective criteria, compared to previous literature. To do so, we depict in Figure \ref{fig:overlap} the distribution of both populations for the proposed methods, AMCons and GradCAMCons, and the most promising baselines, F-anoGAN and Histeq. Furthermore, we obtain the overlap between both distributions by dividing the number of samples in the overlapped region of the histograms by the total number of samples. It can be seen how the proposed method based on entropy maximization obtains the smallest overlap ($10.2\%$) and produces a narrower distribution of normal samples in comparison with the GradCAMCons method, based on size constraints.

\vspace{1 em}
\paragraph{\textbf{Using statistics from normal domain for anomaly localization threshold}} A common practice on unsupervised anomaly segmentation is to use anomalous images to define the threshold to obtain the final segmentation masks. In particular, these methods look at the AUPRC on the anomalous images, which is then used to compute the optimal threshold value. We refer to this technique in our experiments as OP (Operative Point). To alleviate the need of anomalous samples during the validation stage, several methods (\cite{Baur2019DeepImages}) have discussed the possibility of using a given percentile from the normal images (i.e., no anomalies) distribution to set the threshold. Motivated by this, an ablation study on the percentile value is presented in Table \ref{ablationpercentille} for our proposed formulations and the best performing baselines. First, we can observe that under the OP strategy (i.e., accessing to anomalous images to identify the optimal threshold), both of our models bring substantial improvements over the state-of-the-art on residual-based approaches, ranging from 14\% to 22\%. If we resort to the percentiles instead, the performance improvements observed are very similar to the OP scenario, with our models outperforming F-anoGAN by a large margin. Nevertheless, we observed that the best results are obtained with different percentile values. While F-anoGAN and AMCons w. $\mathcal{L}_H$ yields the best performance using the $98\%$ percentile, GradCAMCons w. $\mathcal{L}_S$ follows previous observations in \cite{Baur2019DeepImages}, performing better using the $95\%$ percentile. 

\begin{table}[h!]
\centering
\resizebox{\linewidth}{!}{
\begin{tabular}{|l|c|c|c|c|c|c|}
\hline
         & OP & th=0.5 & p85   & p90    & p95 & p98 \\ \hline\hline
F-anoGAN & $0.525$  & $-$ &  $0.310$ & $0.390$ & $0.505$ & $0.488$ \\ \hline
HistEq      & $0.690$  & $-$ &  $0.298$ & $0.404$ & $0.624$ & $0.620$ \\ \hline
GradCAMCons w. $\mathcal{L}_S$ & $0.693$ & $0.583$ &  $0.512$ & $0.611$ & $0.663$ & $0.587$ \\ \hline
AMCons w. $\mathcal{L}_H$  & $0.743$ & $-$ &  $0.189$ & $0.201$ & $0.265$ & $0.720$ \\ \hline
\end{tabular}
}
\caption{Ablation study on threshold values from normal images. p$X$ indicates the average percentile used on the training set (normal images) to compute the segmentation threshold. OP indicates the operative point from area under precision-recall curve, using all validation dataset, which contains anomalous images. The metric presented is the dataset-level DICE.}
\label{ablationpercentille}
\end{table}

\begin{table*}[h!]
\resizebox{\textwidth}{!}{%
\centering
\small
\begin{tabular}{|l|c|c|c|c|c|}
\hline
\multicolumn{1}{|c|}{\textbf{Method}} &  \multicolumn{1}{c|}{\textbf{AUROC}} & \multicolumn{1}{c|}{\textbf{AUPRC}} & \multicolumn{1}{c|}{\textbf{$\lceil$DICE$\rceil$}} & \multicolumn{1}{c|}{\textbf{$\lceil$IoU$\rceil$}} & \multicolumn{1}{c|}{\textbf{DICE ($\mu\pm\sigma$)}} \\ \hline
\multicolumn{6}{|l|}{\textbf{Other works}} \\ \hline
Karkkainen et al. (2021) (Unsupervised)* & $-$ & $-$ & $-$ & $-$ & $0.197\pm0.222$ \\ \hline
Hssayeni et al. (2020) (Supervised) & $-$ & $-$ & $-$ & $-$ & $0.315\pm0.211$ \\
\hline
\multicolumn{6}{|l|}{\textbf{Physionet-ICH dataset}} \\ \hline
CAVGA (\cite{Venkataramanan2020AttentionImages})   & $0.919$($0.004$) & $0.061$($0.003$) & $0.094$($0.005$) & $0.062$($0.004$) & $0.053$($0.004$)$\pm0.161$($0.002$) \\ \hline
Grad-CAM$_{D}$ VAE (\cite{Liu2020TowardsAutoencoders})  & $0.955$($0.003$) & $0.157$($0.009$) & $0.275$($0.011$) & $0.159$($0.005$) & $0.178$($0.005$)$\pm0.175$($0.003$) \\ \hline
Bayesian AE (\cite{NickPawlowski2018UnsupervisedAutoencoders}) & $0.961$($0.001$) & $0.188$($0.006$) & $0.309$($0.009$) & $0.183$($0.007$) & $0.242$($0.008$)$\pm0.181$($0.003$)\\ \hline
VAE (\cite{Baur2019DeepImages,Zimmerer2020Abstract:Auto-encoders}) & $0.962$($0.000$) & $0.167$($0.005$) & $0.319$($0.002$) & $0.190$($0.002$) & $0.245$($0.004$)$\pm0.192$($0.003$) \\ \hline
AnoVAEGAN (\cite{Baur2019DeepImages}) & $0.961$($0.000$) & $0.167$($0.003$) & $0.313$($0.006$) & $0.185$($0.004$) & $0.239$($0.006$)$\pm0.192$($0.002$) \\ \hline
Bayesian VAE (\cite{NickPawlowski2018UnsupervisedAutoencoders}) & $0.964$($0.000$) & $0.178$($0.010$) & $0.323$($0.007$) & $0.193$($0.005$) & $0.248$($0.008$)$\pm0.191$($0.004$) \\ \hline
Context VAE (\cite{zimmerer2019context}) & $0.963$($0.002$) & $0.170$($0.013$) & $0.321$($0.023$) & $0.191$($0.016$) & $0.243$($0.014$)$\pm0.191$($0.009$) \\ \hline
Restoration VAE (\cite{Chen2020UnsupervisedPrior}) & $0.962$($0.001$) & $0.183$($0.005$) & $0.327$($0.002$) & $0.187$($0.001$) & $0.233$($0.004$)$\pm0.189$($0.003$) \\ \hline
Context AE (\cite{zimmerer2019context}) & $0.962$($0.001$) & $0.195$($0.005$) & $0.359$($0.010$) & $0.219$($0.007$) & $0.276$($0.004$)$\pm0.198$($0.004$) \\ \hline
F-anoGAN (\cite{Schlegl2019F-AnoGAN:Networks}) & $0.961$($0.000$) & $0.173$($0.007$) & $0.343$($0.007$) & $0.207$($0.005$) & $0.268$($0.007$)$\pm0.191$($0.005$) \\ \hline
AE & $0.961$($0.001$) & $0.176$($0.006$) & $0.344$($0.007$) & $0.208$($0.006$) & $0.266$($0.002$)$\pm0.202$($0.005$) \\ \hline
\rowcolor{gray!5} GradCAMCons w. $\mathcal{L}_S$ (L2 penalty) & $0.967$($0.009$) & $0.261$($0.013$) & $0.361$($0.067$) & $0.231$($0.029$) & $0.276$($0.046$)$\pm0.243$($0.029$) \\ \hline
HistEq (\cite{MeissenHistogram}) & $0.963$($0.000$) & $0.313$($0.000$) & $0.385$($0.000$) & $0.239$($0.000$) & $\mathbf{0.348}$($0.000$)$\mathbf{\pm0.213}$($0.000$) \\ \hline
\rowcolor{gray!5} GradCAMCons w. $\mathcal{L}_S$ (Log Barrier) & $0.970$($0.008$) & $0.295$($0.073$) & $0.401$($0.044$) & $0.251$($0.049$) & $0.286$($0.076$)$\pm0.233$($0.039$) \\ \hline
\rowcolor{gray!5}\textbf{AMCons w. $\mathcal{L}_H$} & $\mathbf{0.971}$($0.006$) & $\mathbf{0.420}$($0.068$) & $\mathbf{0.522}$($0.046$) & $\mathbf{0.354}$($0.043$) & $0.319$($0.054$)$\pm0.266$($0.011$) \\ \hline
\multicolumn{6}{l}{\scriptsize{* Results reported on a different (private) dataset}.}

\end{tabular}
}
\caption{Comparison to prior literature on Physionet-ICH dataset, and previous works on ICH segmentation.  Results derived from the proposed methods are depicted in gray, and best results are indicated in bold. The values in parentheses indicate the standard deviation over the three training repetitions.}
\label{tab:test_results_ct}
\vspace{-5mm}
\end{table*}

This suggests that, even though not used directly, anomalous images are still required to find the optimal threshold value. However, the proposed method GradCAMCons shows special properties that suggest that they can achieve large performance gains without having access to anomalous images to define the threshold, unlike prior works. In particular, our GradCAM-based formulation restricts the attention values to $[0, 1]$, which allows to set a typical threshold to $0.5$, with still large performance gains (+7\%) compared to the baselines. Nevertheless, we can observe that if we resort to the percentile strategy, our method based on maximizing the entropy of the attention maps (i.e., AMCons) is very sensitive to the selected value.

\begin{figure}[h!]
    \begin{center}

          \subfloat[\label{fig:ablation_slices_1}]{\includegraphics[width=0.85\linewidth]{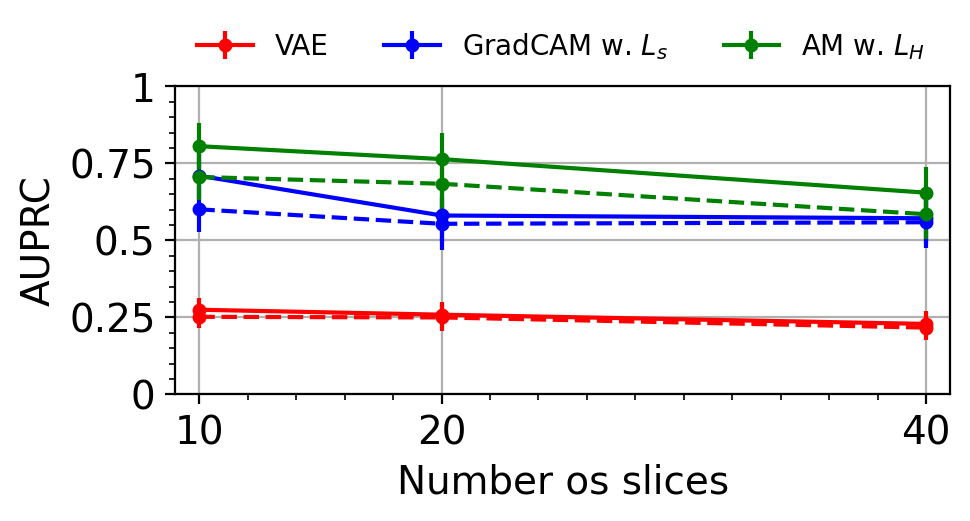}}

          \vspace{-1em}
          \subfloat[\label{fig:ablation_slices_2}]{\includegraphics[width=0.85\linewidth]{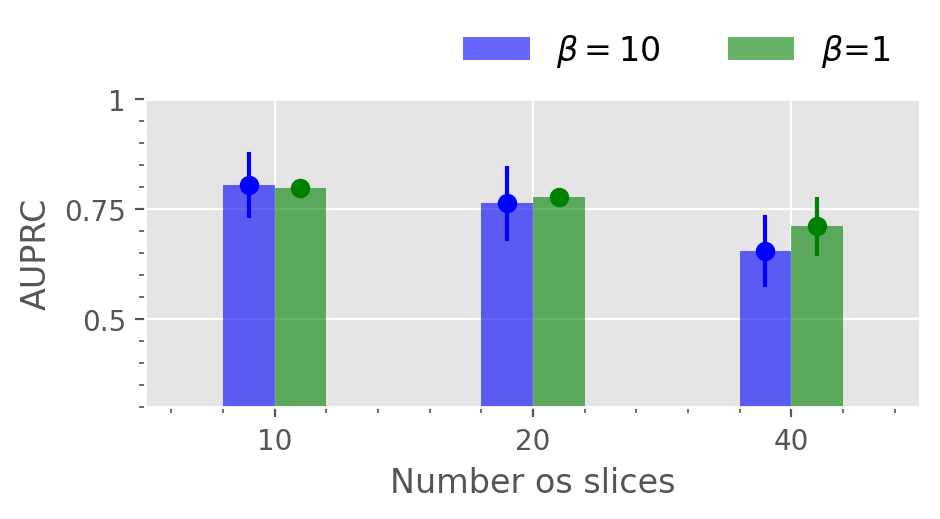}}

        \caption{Ablation study on the effect of increasing the number of axial slices around the center used from MR brain volumes. (a) Study of latent space dimension for the proposed models and an standard VAE. Solid lines indicate $\zz=32$, and dashed lines denote $\zz=128$. (b) Study of the KL component importance ($\beta$ term) using the proposed AMCons method.}
        \label{fig:ablation_slices}
    \end{center}
    \vspace{-2.15 em}
\end{figure}

\paragraph{\textbf{Number of slices to generate the pseudo-volumes}} In our experiments, we followed the standard literature (\cite{Baur2021AutoencodersStudy}) to generate the pseudo-labels for validation and testing. Nevertheless, we concede that this scenario is unrealistic, as the appropriate number of slices used from the MRI scans in unsupervised anomaly detection should be unknown. We now explore the impact of including more slices in these pseudo-volumes, which increase the variability of normal samples. For instance, it is well-known that the target regions in slices farther from the center are incrementally smaller. In this line, we hypothesize that the dimension of the VAE latent space and the importance of the KL regularization may be a determining factors in absorbing this increased variability. Regarding the latent space, the appropriate $\zz$ dimension is unclear in the literature. For instance, \cite{Baur2021AutoencodersStudy} uses $\zz=128$, while \cite{Baur2019DeepImages} uses $\zz=64$, and we obtained better results using $\zz=32$. To validate the proposed experimental setting and latent space dimension, we now present results using increasing number of slices around the axial midline $N=\{10, 20, 40\}$, and two different latent space dimensions $\zz=\{32, 128\}$ for both a standard VAE and our proposed models, in Figure \ref{fig:ablation_slices_1}. We can observe that despite the gap between the baselines and the attention based methods is reduced as the number of slides is increased, this difference is still significant, and the relative performance drop is similar for all methods. Finally, we can observe that an increasing on $\zz$ dimension (solid \textit{versus} dotted lines in Fig \ref{fig:ablation_slices_1}) does not produce gains in performance in any case. Note that the model hyperparameters used are optimized for $z=32$, and $N=10$, which also could produce some underestimation of the proposed model performance when $N$ increases. %
In the following, we study the performance of the proposed AMCons method using different $\beta$ values ($\beta=\{1, 10\}$) in the KL term of eq. \ref{eq:VAE_vanilla} across different number of slices, whose results are presented in Figure \ref{fig:ablation_slices_2}. We can observe that, by decreasing the value of $\beta$ as the number of employed slices increases, we can alleviate the performance degradation observed with a fixed $\beta$. Since the KL regularization directly affects the capacity of the VAE for learning different samples, the optimization of its balancing term when increasing the domain of samples used seems necessary.
The similar behaviour between the proposed method and baselines suggest that this could be a limitation of self-training features based on VAEs, which struggle to encode heterogeneous sample information.

\subsection{Generalization to other datasets}

In order to empirically demonstrate the generalization properties of the proposed methodology, we evaluate its performance on a different dataset for brain lesion detection. Concretely, as previously described, we resort to Physionet-ICH dataset for non-contrast CT on ICH localization. Implementation details are analogous as the ones used on the BraTS dataset, although we decreased the learning rate to $1e-5$, and we set a larger latent dimension, i.e. $\zz\in \real^{128}$, along all baselines and methods to favour model convergence. Obtained results for anomaly localization are reported in Table \ref{tab:test_results_ct}. Even though there exist slight differences in the comparison between residual methods in the literature compared to the results obtained on BraTS dataset (i.e. the simple AE outperforms variations approaches), the proposed attention-based anomaly localization methods still achieve remarkable results. Again, the AMCons configuration yields the best performance, and it reaches improvements of nearly $\sim$25\% and $\sim$18\% in terms of AUPRC and DICE, respectively, compared to previous literature. The observed results suggest that the proposed methodology is able to generalize to other unsupervised brain lesion segmentation challenges, even using different imaging modalities. It should be noted, however, that the absolute results in terms of segmentation are lower than those obtained in BraTS. Among other reasons, this may be due to the greater heterogeneity observed in the ICH dataset, the lower degree of standardization and size of the database used, and the small size of ICH lesions, which penalizes metrics such as DICE. Nevertheless, the values obtained are in line with the scarce previous literature on ICH segmentation, as reflected in Table \ref{tab:test_results_ct}. Indeed, the obtained results are at par with previous works using a fully supervised learning approach \cite{Hssayeni2020IntracranialModel}, which shows the difficulty of the task.

\subsection{Qualitative evaluation}
\begin{figure}[h!]
\begin{center}
\centering
\includegraphics[width=1\linewidth]{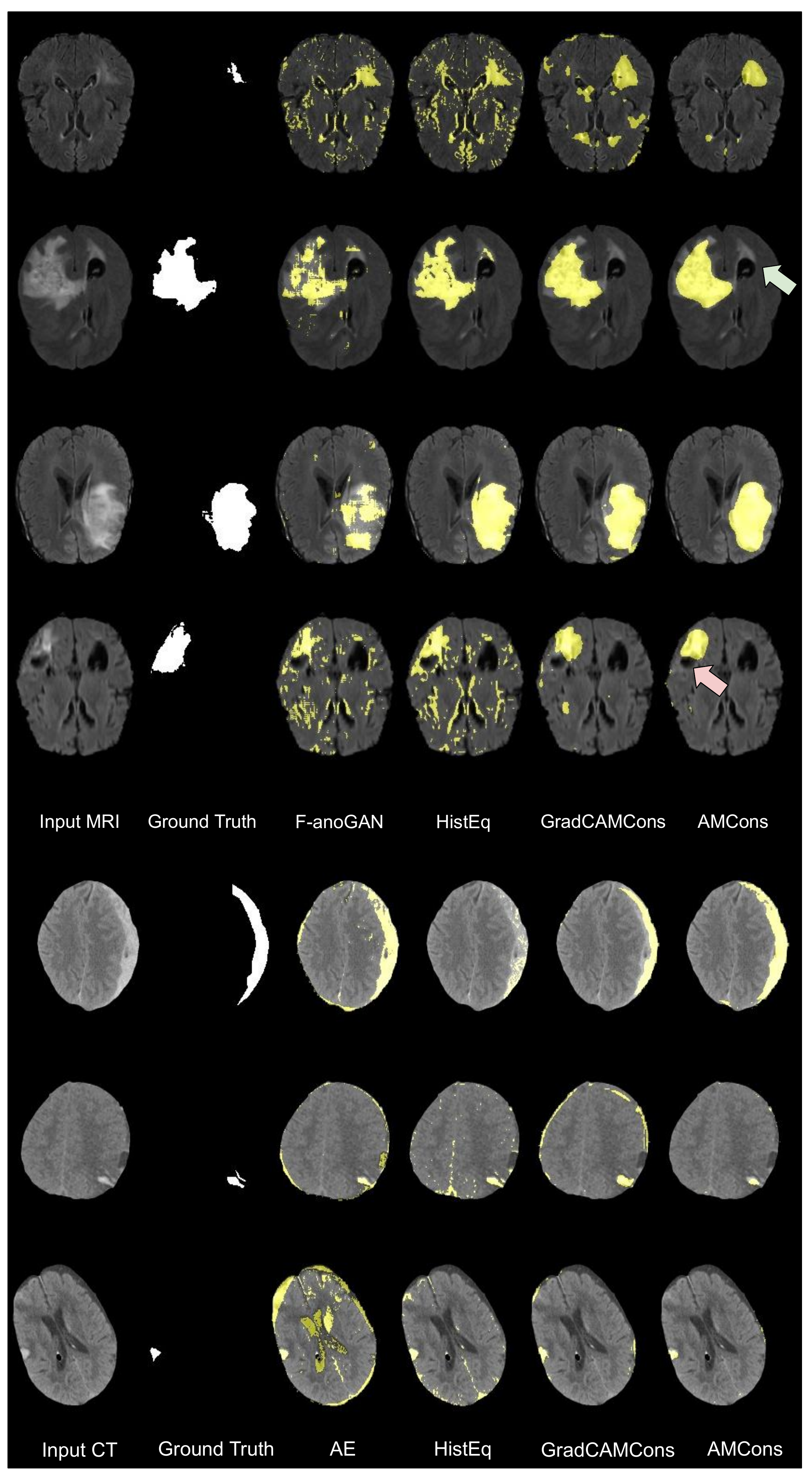} 
\caption{Qualitative evaluation of the proposed and existing high-performing methods for anomaly localization on BraTS MRI flair volumes (top) and on Physionet-ICH non-contrast CT images (bottom). A failure case is depicted with the red arrow (\textit{fourth column}).}
\label{fig:qualitative}
\end{center}
\vspace{-2 em}
\end{figure}

Visual results of the proposed and existing methods for both datasets are depicted in Figure \ref{fig:qualitative}. We can observe that our approach identifies as anomalous more complete regions of the lesions, whereas existing methods are prone to produce a significant amount of false positives (\textit{first, third and seventh} rows) and fail to discover many abnormal pixels (\textit{third row}). These visual results are in line with the quantitative validation performed in previous sections. However, there is a known problem about segmenting only hyperintense regions in the state-of-the-art methods of unsupervised anomaly localization of brain lesions (\cite{MeissenHistogram}). Although the proposed method still suffers from this limitation (\textit{fourth row, red arrow}), the positive results regarding true negative segmentation obtained in some normal, hyperintense tissue (\textit{second row, green arrow}) suggest an improvement in relation to this problem.

\section{Discussion}
\label{sec:conclusions}

Despite the recent advances of unsupervised anomaly segmentation in medical problems, existing literature still provides limited performance, with most methods yielding suboptimal results in popular segmentation benchmarks. In this work, we have presented a novel approach that substantially differs from prior literature in several aspects.

First, we resort to generated attention maps to identify anomalous regions, which contrasts with most existing works that rely on the pixel-wise reconstruction error. Second, our formulation integrates a size-constrained loss that enforces the attention maps to cover the whole image in normal images. This differs from very recent works \cite{Venkataramanan2020AttentionImages}, as we tackle this problem by imposing inequality constraints on the whole target attention maps. 
Another important difference lies on the manner the constrained problem is addressed. While \cite{Venkataramanan2020AttentionImages} leverages a L2 penalty function, we resort to an extension of standard log-barrier methods, which overcome the well-known limitations of penalty-based methods. Quantitative results demonstrate that this model significantly outperforms prior literature on unsupervised lesion segmentation. 

A drawback of the log-barrier based formulation is that it requires to find the optimal value for several hyperparameters. Motivated by this, we have proposed an alternative model, which integrates a regularization term that maximizes the Shannon entropy on the generated attention maps. This new formulation only adds the entropy balancing term $\mathcal{L}_H$, which reduces the complexity compared to the constrained problem in eq. \ref{eq:criterion}. Furthermore, as reported in the results, the maximum-entropy model yields better performance than the size regularizer formulation. Note, in addition, that the alternative entropy-based model better separates the intensity distributions between normal and abnormal tissue. This allows us to employ a higher percentile value to obtain the final anomalous regions, with a substantial performance improvement compared to previous methods. Thus, based on the reported empirical validation, the proposed models represent a novel state-of-the-art for unsupervised anomaly segmentation.

We believe that there exist potential research directions to further improve the performance of unsupervised segmentation methods. For example, brain images are typically acquired along multiple modalities. Learning how to combine multiple modalities in the scenario of anomalous regions detection might indeed enhance the learned representation by the VAE, ultimately resulting in better identification of abnormal pixels. In addition, unsupervised segmentation methods have been only evaluated from a discriminative perspective. Nevertheless, assessing their performances in terms of the quality of the uncertainty estimates, i.e., calibration, might give a better overview of the quality of a segmentation model.

\section*{Acknowledgments}

J.~Silva-Rodr\'iguez work was supported by the Spanish Government under FPI Grant PRE2018-083443. The DGX-A100 used in this work was partially funded by Generalitat Valenciana / European Union through the European Regional Development Fund (ERDF) of the Valencian Community (IDIFEDER/2020/030).

\bibliographystyle{model2-names.bst}\biboptions{authoryear}
\bibliography{refs}

\begin{thebibliography}{52}
\expandafter\ifx\csname natexlab\endcsname\relax\def\natexlab#1{#1}\fi
\providecommand{\url}[1]{\texttt{#1}}
\providecommand{\href}[2]{#2}
\providecommand{\path}[1]{#1}
\providecommand{\DOIprefix}{doi:}
\providecommand{\ArXivprefix}{arXiv:}
\providecommand{\URLprefix}{URL: }
\providecommand{\Pubmedprefix}{pmid:}
\providecommand{\doi}[1]{\href{http://dx.doi.org/#1}{\path{#1}}}
\providecommand{\Pubmed}[1]{\href{pmid:#1}{\path{#1}}}
\providecommand{\bibinfo}[2]{#2}
\ifx\xfnm\relax \def\xfnm[#1]{\unskip,\space#1}\fi
\bibitem[{Abati et~al.(2019)Abati, Porrello, Calderara and
  Cucchiara}]{Abati2019LatentDetection}
\bibinfo{author}{Abati, D.}, \bibinfo{author}{Porrello, A.},
  \bibinfo{author}{Calderara, S.}, \bibinfo{author}{Cucchiara, R.},
  \bibinfo{year}{2019}.
\newblock \bibinfo{title}{{Latent space autoregression for novelty detection}},
  in: \bibinfo{booktitle}{Proceedings of the IEEE Computer Society Conference
  on Computer Vision and Pattern Recognition (CVPR)}.
\bibitem[{Andermatt et~al.(2019)Andermatt, Horv{\'{a}}th, Pezold and
  Cattin}]{Andermatt2019PathologyOrigin}
\bibinfo{author}{Andermatt, S.}, \bibinfo{author}{Horv{\'{a}}th, A.},
  \bibinfo{author}{Pezold, S.}, \bibinfo{author}{Cattin, P.},
  \bibinfo{year}{2019}.
\newblock \bibinfo{title}{{Pathology segmentation using distributional
  differences to images of healthy origin}}.
\newblock \bibinfo{journal}{Medical Image Computing and Computer Assisted
  Intervention (MICCAI) - Brainlesion Workshop} .
\bibitem[{Bakas et~al.(2017)Bakas, Akbari, Sotiras, Bilello, Rozycki, Kirby,
  Freymann, Farahani and Davatzikos}]{Bakas2017AdvancingFeaturesJ}
\bibinfo{author}{Bakas, S.}, \bibinfo{author}{Akbari, H.},
  \bibinfo{author}{Sotiras, A.}, \bibinfo{author}{Bilello, M.},
  \bibinfo{author}{Rozycki, M.}, \bibinfo{author}{Kirby, J.S.},
  \bibinfo{author}{Freymann, J.B.}, \bibinfo{author}{Farahani, K.},
  \bibinfo{author}{Davatzikos, C.}, \bibinfo{year}{2017}.
\newblock \bibinfo{title}{{Advancing The Cancer Genome Atlas glioma MRI
  collections with expert segmentation labels and radiomic features}}.
\newblock \bibinfo{journal}{Scientific Data} \bibinfo{volume}{4},
  \bibinfo{pages}{1--13}.
\bibitem[{Bakas et~al.(2018)}]{Bakas2018IdentifyingChallengeJ}
\bibinfo{author}{Bakas, S.}, et~al., \bibinfo{year}{2018}.
\newblock \bibinfo{title}{{Identifying the Best Machine Learning Algorithms for
  Brain Tumor Segmentation, Progression Assessment, and Overall Survival
  Prediction in the BRATS Challenge}} .
\bibitem[{Bateson et~al.(2021)Bateson, Dolz, Kervadec, Lombaert and
  Ayed}]{Bateson2021ConstrainedSegmentation}
\bibinfo{author}{Bateson, M.}, \bibinfo{author}{Dolz, J.},
  \bibinfo{author}{Kervadec, H.}, \bibinfo{author}{Lombaert, H.},
  \bibinfo{author}{Ayed, I.B.}, \bibinfo{year}{2021}.
\newblock \bibinfo{title}{{Constrained Domain Adaptation for Image
  Segmentation}}.
\newblock \bibinfo{journal}{IEEE Transactions on Medical Imaging}
  \bibinfo{volume}{40}, \bibinfo{pages}{1875--1887}.
\bibitem[{Baur et~al.(2021)Baur, Denner, Wiestler, Navab and
  Albarqouni}]{Baur2021AutoencodersStudy}
\bibinfo{author}{Baur, C.}, \bibinfo{author}{Denner, S.},
  \bibinfo{author}{Wiestler, B.}, \bibinfo{author}{Navab, N.},
  \bibinfo{author}{Albarqouni, S.}, \bibinfo{year}{2021}.
\newblock \bibinfo{title}{{Autoencoders for unsupervised anomaly segmentation
  in brain MR images: A comparative study}}.
\newblock \bibinfo{journal}{Medical Image Analysis} \bibinfo{volume}{69},
  \bibinfo{pages}{1--16}.
\bibitem[{Baur et~al.(2020)Baur, Graf, Wiestler, Albarqouni and
  Navab}]{Baur2020SteGANomaly:MRI}
\bibinfo{author}{Baur, C.}, \bibinfo{author}{Graf, R.},
  \bibinfo{author}{Wiestler, B.}, \bibinfo{author}{Albarqouni, S.},
  \bibinfo{author}{Navab, N.}, \bibinfo{year}{2020}.
\newblock \bibinfo{title}{{SteGANomaly: Inhibiting CycleGAN Steganography for
  Unsupervised Anomaly Detection in Brain MRI}}, in:
  \bibinfo{booktitle}{Medical Image Computing and Computer Assisted
  Intervention (MICCAI)}, \bibinfo{publisher}{Springer International
  Publishing}.
\bibitem[{Baur et~al.(2019)Baur, Wiestler, Albarqouni and
  Navab}]{Baur2019DeepImages}
\bibinfo{author}{Baur, C.}, \bibinfo{author}{Wiestler, B.},
  \bibinfo{author}{Albarqouni, S.}, \bibinfo{author}{Navab, N.},
  \bibinfo{year}{2019}.
\newblock \bibinfo{title}{{Deep autoencoding models for unsupervised anomaly
  segmentation in brain MR images}}, in: \bibinfo{booktitle}{Medical Image
  Computing and Computer Assisted Intervention (MICCAI)}.
\bibitem[{Bergmann et~al.(2020)Bergmann, Fauser, Sattlegger and
  Steger}]{bergmann2020uninformed}
\bibinfo{author}{Bergmann, P.}, \bibinfo{author}{Fauser, M.},
  \bibinfo{author}{Sattlegger, D.}, \bibinfo{author}{Steger, C.},
  \bibinfo{year}{2020}.
\newblock \bibinfo{title}{Uninformed students: Student-teacher anomaly
  detection with discriminative latent embeddings}, in:
  \bibinfo{booktitle}{Proceedings of the IEEE/CVF Conference on Computer Vision
  and Pattern Recognition}, pp. \bibinfo{pages}{4183--4192}.
\bibitem[{Bergmann et~al.(2019)Bergmann, L{\"{o}}we, Fauser, Sattlegger and
  Steger}]{Bergmann2019ImprovingAutoencoders}
\bibinfo{author}{Bergmann, P.}, \bibinfo{author}{L{\"{o}}we, S.},
  \bibinfo{author}{Fauser, M.}, \bibinfo{author}{Sattlegger, D.},
  \bibinfo{author}{Steger, C.}, \bibinfo{year}{2019}.
\newblock \bibinfo{title}{{Improving unsupervised defect segmentation by
  applying structural similarity to autoencoders}}, in:
  \bibinfo{booktitle}{Proceedings of the 14th International Joint Conference on
  Computer Vision, Imaging and Computer Graphics Theory and Applications
  (VISIGRAPP )}.
\bibitem[{Boyd et~al.(2004)Boyd, Boyd and Vandenberghe}]{boyd2004convex}
\bibinfo{author}{Boyd, S.}, \bibinfo{author}{Boyd, S.P.},
  \bibinfo{author}{Vandenberghe, L.}, \bibinfo{year}{2004}.
\newblock \bibinfo{title}{Convex optimization}.
\newblock \bibinfo{publisher}{Cambridge university press}.
\bibitem[{Chen and Konukoglu(2018)}]{Chen2018UnsupervisedAuto-encoders}
\bibinfo{author}{Chen, X.}, \bibinfo{author}{Konukoglu, E.},
  \bibinfo{year}{2018}.
\newblock \bibinfo{title}{{Unsupervised Detection of Lesions in Brain MRI using
  constrained adversarial auto-encoders}}, in: \bibinfo{booktitle}{Medical
  Imaging with Deep Learning (MIDL)}.
\bibitem[{Chen et~al.(2020)Chen, You, Tezcan and
  Konukoglu}]{Chen2020UnsupervisedPrior}
\bibinfo{author}{Chen, X.}, \bibinfo{author}{You, S.}, \bibinfo{author}{Tezcan,
  K.C.}, \bibinfo{author}{Konukoglu, E.}, \bibinfo{year}{2020}.
\newblock \bibinfo{title}{{Unsupervised lesion detection via image restoration
  with a normative prior}}.
\newblock \bibinfo{journal}{Medical Image Analysis} \bibinfo{volume}{64}.
\bibitem[{Dehaene et~al.(2020)Dehaene, Frigo, Combrexelle and
  Eline}]{Dehaene2020IterativeLocalization}
\bibinfo{author}{Dehaene, D.}, \bibinfo{author}{Frigo, O.},
  \bibinfo{author}{Combrexelle, S.}, \bibinfo{author}{Eline, P.},
  \bibinfo{year}{2020}.
\newblock \bibinfo{title}{{Iterative energy-based projection on a normal data
  manifold for anomaly localization}}, in: \bibinfo{booktitle}{Proceedings of
  the International Conference on Learning Representations (ICLR)}.
\bibitem[{Fiacco and
  McCormick(1990)}]{FiaccoAnthonyVandMcCormick1990NonlinearTechniques}
\bibinfo{author}{Fiacco, A.V.}, \bibinfo{author}{McCormick, G.P.},
  \bibinfo{year}{1990}.
\newblock \bibinfo{title}{{Nonlinear programming: sequential unconstrained
  minimization techniques}}.
\newblock \bibinfo{publisher}{SIAM}.
\bibitem[{Goldberger et~al.(2000)Goldberger, Amaral, Glass, Hausdorff, Ivanov,
  Mark, Mietus, Moody, Peng and Stanley}]{Goldberger2000PhysioBankSignals_blue}
\bibinfo{author}{Goldberger, A.L.}, \bibinfo{author}{Amaral, L.A.N.},
  \bibinfo{author}{Glass, L.}, \bibinfo{author}{Hausdorff, J.M.},
  \bibinfo{author}{Ivanov, P.C.}, \bibinfo{author}{Mark, R.G.},
  \bibinfo{author}{Mietus, J.E.}, \bibinfo{author}{Moody, G.B.},
  \bibinfo{author}{Peng, C.k.}, \bibinfo{author}{Stanley, H.E.},
  \bibinfo{year}{2000}.
\newblock \bibinfo{title}{{PhysioBank, PhysionToolkit, and PhysioNet:
  Components of a new research resource for complex physiologic signals}}.
\newblock \bibinfo{journal}{Circulation} \bibinfo{volume}{101}.
\bibitem[{Goodfellow et~al.(2014)Goodfellow, Pouget-Abadie, Mirza, Xu,
  Warde-Farley, Ozair, Courville and Bengio}]{Goodfellow2014GenerativeNetworks}
\bibinfo{author}{Goodfellow, I.}, \bibinfo{author}{Pouget-Abadie, J.},
  \bibinfo{author}{Mirza, M.}, \bibinfo{author}{Xu, B.},
  \bibinfo{author}{Warde-Farley, D.}, \bibinfo{author}{Ozair, S.},
  \bibinfo{author}{Courville, A.}, \bibinfo{author}{Bengio, Y.},
  \bibinfo{year}{2014}.
\newblock \bibinfo{title}{{Generative adversarial networks}}, in:
  \bibinfo{booktitle}{Advances in Neural Information Processing Systems
  (NeurIPS)}.
\bibitem[{He et~al.(2017)He, Liu, Schwing and Peng}]{He2017LearningTightening}
\bibinfo{author}{He, F.S.}, \bibinfo{author}{Liu, Y.},
  \bibinfo{author}{Schwing, A.G.}, \bibinfo{author}{Peng, J.},
  \bibinfo{year}{2017}.
\newblock \bibinfo{title}{{Learning to play in a day: Faster deep reinforcement
  learning by optimality tightening}}, in: \bibinfo{booktitle}{Proceedings of
  the International Conference on Learning Representations (ICLR)}, pp.
  \bibinfo{pages}{1--13}.
\bibitem[{He et~al.(2016)He, Zhang, Ren and Sun}]{He2016DeepRecognition}
\bibinfo{author}{He, K.}, \bibinfo{author}{Zhang, X.}, \bibinfo{author}{Ren,
  S.}, \bibinfo{author}{Sun, J.}, \bibinfo{year}{2016}.
\newblock \bibinfo{title}{{Deep residual learning for image recognition}}, in:
  \bibinfo{booktitle}{Proceedings of the Conference on Computer Vision and
  Pattern Recognition (CVPR)}.
\bibitem[{Hssayeni(2020)}]{Hssayeni2020ComputedSegmentation_blue}
\bibinfo{author}{Hssayeni, M.}, \bibinfo{year}{2020}.
\newblock \bibinfo{title}{{Computed Tomography Images for Intracranial
  Hemorrhage Detection and Segmentation}}.
\newblock \bibinfo{journal}{PhysioNet} .
\bibitem[{Hssayeni et~al.(2020a)Hssayeni, Croock, Salman, Al-Khafaji, Yahya and
  Ghoraani}]{Hssayeni2020IntracranialModel_blue}
\bibinfo{author}{Hssayeni, M.D.}, \bibinfo{author}{Croock, M.S.},
  \bibinfo{author}{Salman, A.D.}, \bibinfo{author}{Al-Khafaji, H.F.},
  \bibinfo{author}{Yahya, Z.A.}, \bibinfo{author}{Ghoraani, B.},
  \bibinfo{year}{2020}a.
\newblock \bibinfo{title}{{Intracranial hemorrhage segmentation using a deep
  convolutional model}}.
\newblock \bibinfo{journal}{Data} \bibinfo{volume}{5}, \bibinfo{pages}{1--18}.
\bibitem[{Hssayeni et~al.(2020b)Hssayeni, Croock, Salman, Al-Khafaji, Yahya and
  Ghoraani}]{Hssayeni2020IntracranialModel}
\bibinfo{author}{Hssayeni, M.D.}, \bibinfo{author}{Croock, M.S.},
  \bibinfo{author}{Salman, A.D.}, \bibinfo{author}{Al-Khafaji, H.F.},
  \bibinfo{author}{Yahya, Z.A.}, \bibinfo{author}{Ghoraani, B.},
  \bibinfo{year}{2020}b.
\newblock \bibinfo{title}{{Intracranial hemorrhage segmentation using a deep
  convolutional model}}.
\newblock \bibinfo{journal}{Data} \bibinfo{volume}{5}, \bibinfo{pages}{1--18}.
\bibitem[{Ilse et~al.(2018)Ilse, Tomczak and
  Welling}]{Ilse2018Attention-basedLearning}
\bibinfo{author}{Ilse, M.}, \bibinfo{author}{Tomczak, J.M.},
  \bibinfo{author}{Welling, M.}, \bibinfo{year}{2018}.
\newblock \bibinfo{title}{{Attention-based deep multiple instance learning}},
  in: \bibinfo{booktitle}{35th International Conference on Machine Learning
  (ICML)}.
\bibitem[{Jia et~al.(2017)Jia, Huang, Chang and
  Xu}]{Jia2017ConstrainedSegmentation}
\bibinfo{author}{Jia, Z.}, \bibinfo{author}{Huang, X.}, \bibinfo{author}{Chang,
  E.I.}, \bibinfo{author}{Xu, Y.}, \bibinfo{year}{2017}.
\newblock \bibinfo{title}{{Constrained Deep Weak Supervision for Histopathology
  Image Segmentation}}.
\newblock \bibinfo{journal}{IEEE Transactions on Medical Imaging}
  \bibinfo{volume}{36}, \bibinfo{pages}{2376--2388}.
\bibitem[{Kervadec et~al.(2019a)Kervadec, Dolz, Granger and
  Ben~Ayed}]{Kervadec2019CurriculumSegmentation}
\bibinfo{author}{Kervadec, H.}, \bibinfo{author}{Dolz, J.},
  \bibinfo{author}{Granger, E.}, \bibinfo{author}{Ben~Ayed, I.},
  \bibinfo{year}{2019}a.
\newblock \bibinfo{title}{{Curriculum Semi-supervised Segmentation}}, in:
  \bibinfo{booktitle}{Medical Image Computing and Computer Assisted
  Intervention (MICCAI)}.
\bibitem[{Kervadec et~al.(2019b)Kervadec, Dolz, Tang, Granger, Boykov and
  Ben~Ayed}]{Kervadec2019Constrained-CNNSegmentation}
\bibinfo{author}{Kervadec, H.}, \bibinfo{author}{Dolz, J.},
  \bibinfo{author}{Tang, M.}, \bibinfo{author}{Granger, E.},
  \bibinfo{author}{Boykov, Y.}, \bibinfo{author}{Ben~Ayed, I.},
  \bibinfo{year}{2019}b.
\newblock \bibinfo{title}{{Constrained-CNN losses for weakly supervised
  segmentation}}.
\newblock \bibinfo{journal}{Medical Image Analysis} \bibinfo{volume}{54},
  \bibinfo{pages}{88--99}.
\bibitem[{Kervadec et~al.(2019c)Kervadec, Dolz, Yuan, Desrosiers, Granger and
  Ayed}]{Kervadec2019ConstrainedExtensions}
\bibinfo{author}{Kervadec, H.}, \bibinfo{author}{Dolz, J.},
  \bibinfo{author}{Yuan, J.}, \bibinfo{author}{Desrosiers, C.},
  \bibinfo{author}{Granger, E.}, \bibinfo{author}{Ayed, I.B.},
  \bibinfo{year}{2019}c.
\newblock \bibinfo{title}{{Constrained Deep Networks: Lagrangian Optimization
  via Log-Barrier Extensions}}.
\newblock \bibinfo{journal}{ArXiv Preprint} \URLprefix
  \url{http://arxiv.org/abs/1904.04205}.
\bibitem[{Kingma and Welling(2014)}]{Kingma2014Auto-encodingBayes}
\bibinfo{author}{Kingma, D.P.}, \bibinfo{author}{Welling, M.},
  \bibinfo{year}{2014}.
\newblock \bibinfo{title}{{Auto-encoding variational bayes}}, in:
  \bibinfo{booktitle}{2nd International Conference on Learning Representations,
  (ICLR)}.
\bibitem[{Liu et~al.(2020)Liu, Li, Zheng, Karanam, Wu, Bhanu, Radke and
  Camps}]{Liu2020TowardsAutoencoders}
\bibinfo{author}{Liu, W.}, \bibinfo{author}{Li, R.}, \bibinfo{author}{Zheng,
  M.}, \bibinfo{author}{Karanam, S.}, \bibinfo{author}{Wu, Z.},
  \bibinfo{author}{Bhanu, B.}, \bibinfo{author}{Radke, R.J.},
  \bibinfo{author}{Camps, O.}, \bibinfo{year}{2020}.
\newblock \bibinfo{title}{{Towards Visually Explaining Variational
  Autoencoders}}, in: \bibinfo{booktitle}{Proceedings of the IEEE Computer
  Society Conference on Computer Vision and Pattern Recognition (CVPR)}.
\bibitem[{Luenberger(1973)}]{Luenberger1973IntroductionProgramming}
\bibinfo{author}{Luenberger, D.G.}, \bibinfo{year}{1973}.
\newblock \bibinfo{title}{{Introduction to linear and nonlinear programming}}.
\newblock \bibinfo{publisher}{Addison-wesley Reading, MA}.
\bibitem[{Meissen et~al.(2021)Meissen, Georgios and
  Rueckert}]{MeissenHistogram}
\bibinfo{author}{Meissen, F.}, \bibinfo{author}{Georgios, K.},
  \bibinfo{author}{Rueckert, D.}, \bibinfo{year}{2021}.
\newblock \bibinfo{title}{Challenging current semi-supervised anomaly
  segmentation methods for brain mri}, in: \bibinfo{booktitle}{MICCAI 2021
  BrainLes Workshop}.
\bibitem[{Meissen et~al.(2022)Meissen, Weistler, Kaissis and
  Rueckert}]{MeissenHistogram2}
\bibinfo{author}{Meissen, F.}, \bibinfo{author}{Weistler, B.},
  \bibinfo{author}{Kaissis, G.}, \bibinfo{author}{Rueckert, D.},
  \bibinfo{year}{2022}.
\newblock \bibinfo{title}{On the pitfalls of using the residual error as
  anomaly score}, in: \bibinfo{booktitle}{Medical Image with Deep Learning
  (MIDL)}.
\bibitem[{Menze et~al.(2015)}]{Menze2015TheBRATSJ}
\bibinfo{author}{Menze, B.}, et~al., \bibinfo{year}{2015}.
\newblock \bibinfo{title}{{The Multimodal Brain Tumor Image Segmentation
  Benchmark (BRATS)}}.
\newblock \bibinfo{journal}{IEEE Transactions on Medical Imaging}
  \bibinfo{volume}{34}, \bibinfo{pages}{1993--2024}.
\bibitem[{Nguyen et~al.(2021)Nguyen, Feldman, Bethapudi, Jennings and
  Willcocks}]{nguyen2021unsupervised}
\bibinfo{author}{Nguyen, B.}, \bibinfo{author}{Feldman, A.},
  \bibinfo{author}{Bethapudi, S.}, \bibinfo{author}{Jennings, A.},
  \bibinfo{author}{Willcocks, C.G.}, \bibinfo{year}{2021}.
\newblock \bibinfo{title}{Unsupervised region-based anomaly detection in brain
  {MRI} with adversarial image inpainting}, in: \bibinfo{booktitle}{2021 IEEE
  18th International Symposium on Biomedical Imaging (ISBI)},
  \bibinfo{organization}{IEEE}. pp. \bibinfo{pages}{1127--1131}.
\bibitem[{Nick~Pawlowski(2018)}]{NickPawlowski2018UnsupervisedAutoencoders}
\bibinfo{author}{Nick~Pawlowski, M.C.L.}, \bibinfo{year}{2018}.
\newblock \bibinfo{title}{{Unsupervised Lesion Detection in Brain CT using
  Bayesian Convolutional Autoencoders}}, in: \bibinfo{booktitle}{Medical
  Imaging with Deep Learning (MIDL)}.
\bibitem[{Pathak et~al.(2015)Pathak, Krahenbuhl and
  Darrell}]{Pathak2015ConstrainedSegmentation}
\bibinfo{author}{Pathak, D.}, \bibinfo{author}{Krahenbuhl, P.},
  \bibinfo{author}{Darrell, T.}, \bibinfo{year}{2015}.
\newblock \bibinfo{title}{{Constrained convolutional neural networks for weakly
  supervised segmentation}}.
\newblock \bibinfo{journal}{Proceedings of the IEEE International Conference on
  Computer Vision (ICCV)} , \bibinfo{pages}{1796--1804}.
\bibitem[{Peng et~al.(2020)Peng, Kervadec, Dolz, Ben~Ayed, Pedersoli and
  Desrosiers}]{Peng2020Discretely-constrainedSegmentation}
\bibinfo{author}{Peng, J.}, \bibinfo{author}{Kervadec, H.},
  \bibinfo{author}{Dolz, J.}, \bibinfo{author}{Ben~Ayed, I.},
  \bibinfo{author}{Pedersoli, M.}, \bibinfo{author}{Desrosiers, C.},
  \bibinfo{year}{2020}.
\newblock \bibinfo{title}{{Discretely-constrained deep network for weakly
  supervised segmentation}}.
\newblock \bibinfo{journal}{Neural Networks} \bibinfo{volume}{130},
  \bibinfo{pages}{297--308}.
\bibitem[{Radford et~al.(2016)Radford, Metz and
  Chintala}]{Radford2016UnsupervisedNetworks}
\bibinfo{author}{Radford, A.}, \bibinfo{author}{Metz, L.},
  \bibinfo{author}{Chintala, S.}, \bibinfo{year}{2016}.
\newblock \bibinfo{title}{{Unsupervised representation learning with deep
  convolutional generative adversarial networks}}, in:
  \bibinfo{booktitle}{International Conference on Learning Representations
  (ICLR)}.
\bibitem[{Ravanbakhsh et~al.(2019)Ravanbakhsh, Sangineto, Nabi and
  Sebe}]{Ravanbakhsh2019TrainingCrowds}
\bibinfo{author}{Ravanbakhsh, M.}, \bibinfo{author}{Sangineto, E.},
  \bibinfo{author}{Nabi, M.}, \bibinfo{author}{Sebe, N.}, \bibinfo{year}{2019}.
\newblock \bibinfo{title}{{Training adversarial discriminators for
  cross-channel abnormal event detection in crowds}}.
\newblock \bibinfo{journal}{Proceedings of the IEEE Winter Conference on
  Applications of Computer Vision (WACV)} \bibinfo{volume}{2}.
\bibitem[{Sabokrou et~al.(2019)Sabokrou, Pourreza, Fayyaz, Entezari, Fathy,
  Gall and Adeli}]{Sabokrou2019AVID:Detection}
\bibinfo{author}{Sabokrou, M.}, \bibinfo{author}{Pourreza, M.},
  \bibinfo{author}{Fayyaz, M.}, \bibinfo{author}{Entezari, R.},
  \bibinfo{author}{Fathy, M.}, \bibinfo{author}{Gall, J.},
  \bibinfo{author}{Adeli, E.}, \bibinfo{year}{2019}.
\newblock \bibinfo{title}{{AVID: Adversarial Visual Irregularity Detection}},
  in: \bibinfo{booktitle}{Proceedings of the Asia Conference on Computer Vision
  (ACCV)}.
\bibitem[{Schlegl et~al.(2019)Schlegl, Seeb{\"{o}}ck, Waldstein, Langs and
  Schmidt-Erfurth}]{Schlegl2019F-AnoGAN:Networks}
\bibinfo{author}{Schlegl, T.}, \bibinfo{author}{Seeb{\"{o}}ck, P.},
  \bibinfo{author}{Waldstein, S.M.}, \bibinfo{author}{Langs, G.},
  \bibinfo{author}{Schmidt-Erfurth, U.}, \bibinfo{year}{2019}.
\newblock \bibinfo{title}{{f-AnoGAN: Fast unsupervised anomaly detection with
  generative adversarial networks}}.
\newblock \bibinfo{journal}{Medical Image Analysis} \bibinfo{volume}{54},
  \bibinfo{pages}{30--44}.
\bibitem[{Schlegl et~al.(2017)Schlegl, Seeb{\"{o}}ck, Waldstein,
  Schmidt-Erfurth and Langs}]{Schlegl2017UnsupervisedDiscovery}
\bibinfo{author}{Schlegl, T.}, \bibinfo{author}{Seeb{\"{o}}ck, P.},
  \bibinfo{author}{Waldstein, S.M.}, \bibinfo{author}{Schmidt-Erfurth, U.},
  \bibinfo{author}{Langs, G.}, \bibinfo{year}{2017}.
\newblock \bibinfo{title}{{Unsupervised anomaly detection with generative
  adversarial networks to guide marker discovery}}, in:
  \bibinfo{booktitle}{Proceedings of the International Conference on
  Information Processing in Medical Imaging (IPMI)}.
\bibitem[{Selvaraju et~al.(2020)Selvaraju, Cogswell, Das, Vedantam, Parikh and
  Batra}]{Selvaraju2020Grad-CAM:Localization}
\bibinfo{author}{Selvaraju, R.R.}, \bibinfo{author}{Cogswell, M.},
  \bibinfo{author}{Das, A.}, \bibinfo{author}{Vedantam, R.},
  \bibinfo{author}{Parikh, D.}, \bibinfo{author}{Batra, D.},
  \bibinfo{year}{2020}.
\newblock \bibinfo{title}{{Grad-CAM: Visual Explanations from Deep Networks via
  Gradient-Based Localization}}.
\newblock \bibinfo{journal}{International Journal of Computer Vision}
  \bibinfo{volume}{128}, \bibinfo{pages}{336--359}.
\bibitem[{Shi et~al.(2021)Shi, Yang and Qi}]{shi2021unsupervised}
\bibinfo{author}{Shi, Y.}, \bibinfo{author}{Yang, J.}, \bibinfo{author}{Qi,
  Z.}, \bibinfo{year}{2021}.
\newblock \bibinfo{title}{Unsupervised anomaly segmentation via deep feature
  reconstruction}.
\newblock \bibinfo{journal}{Neurocomputing} \bibinfo{volume}{424},
  \bibinfo{pages}{9--22}.
\bibitem[{Silva-Rodr{\'{i}}guez et~al.(2021)Silva-Rodr{\'{i}}guez, Naranjo and
  Dolz}]{Silva-Rodriguez2021LookingSegmentation}
\bibinfo{author}{Silva-Rodr{\'{i}}guez, J.}, \bibinfo{author}{Naranjo, V.},
  \bibinfo{author}{Dolz, J.}, \bibinfo{year}{2021}.
\newblock \bibinfo{title}{{Looking at the whole picture: constrained
  unsupervised anomaly segmentation}}, in: \bibinfo{booktitle}{British Machine
  Vision Conference (BMVC)}.
\bibitem[{Sun et~al.(2020)Sun, Wang, Huang, Ding, Greenspan and
  Paisley}]{Sun2020AnDetection}
\bibinfo{author}{Sun, L.}, \bibinfo{author}{Wang, J.}, \bibinfo{author}{Huang,
  Y.}, \bibinfo{author}{Ding, X.}, \bibinfo{author}{Greenspan, H.},
  \bibinfo{author}{Paisley, J.}, \bibinfo{year}{2020}.
\newblock \bibinfo{title}{{An adversarial learning approach to medical image
  synthesis for lesion detection}}.
\newblock \bibinfo{journal}{IEEE Journal of Biomedical and Health Informatics}
  \bibinfo{volume}{24}, \bibinfo{pages}{2303--2314}.
\bibitem[{Venkataramanan et~al.(2020)Venkataramanan, Peng, Singh and
  Mahalanobis}]{Venkataramanan2020AttentionImages}
\bibinfo{author}{Venkataramanan, S.}, \bibinfo{author}{Peng, K.C.},
  \bibinfo{author}{Singh, R.V.}, \bibinfo{author}{Mahalanobis, A.},
  \bibinfo{year}{2020}.
\newblock \bibinfo{title}{{Attention Guided Anomaly Localization in Images}},
  in: \bibinfo{booktitle}{Proceedings of the European COnference on Computer
  Vision (ECCV)}.
\bibitem[{Zhang et~al.(2017)Zhang, David and Gong}]{Zhang2017CurriculumScenes}
\bibinfo{author}{Zhang, Y.}, \bibinfo{author}{David, P.},
  \bibinfo{author}{Gong, B.}, \bibinfo{year}{2017}.
\newblock \bibinfo{title}{{Curriculum Domain Adaptation for Semantic
  Segmentation of Urban Scenes}}, in: \bibinfo{booktitle}{Proceedings of the
  IEEE International Conference on Computer Vision (ICCV)}.
\bibitem[{Zhou et~al.(2019)Zhou, Li, Bai, Chen, Han, Wang, Fishman and
  Yuille}]{Zhou2019Prior-awareSegmentation}
\bibinfo{author}{Zhou, Y.}, \bibinfo{author}{Li, Z.}, \bibinfo{author}{Bai,
  S.}, \bibinfo{author}{Chen, X.}, \bibinfo{author}{Han, M.},
  \bibinfo{author}{Wang, C.}, \bibinfo{author}{Fishman, E.},
  \bibinfo{author}{Yuille, A.}, \bibinfo{year}{2019}.
\newblock \bibinfo{title}{{Prior-aware neural network for partially-supervised
  multi-organ segmentation}}, in: \bibinfo{booktitle}{Proceedings of the IEEE
  International Conference on Computer Vision (ICCV)}.
\bibitem[{Zimmerer et~al.(2022)Zimmerer, Full, Isensee, Jager, Adler, Petersen,
  Kohler, Ross, Reinke, Kascenas, Jensen, O'Neil, Tan, Hou, Batten, Qiu, Kainz,
  Shvetsova, Fedulova, Dylov, Yu, Zhai, Hu, Si, Zhou, Wang, Li, Chen, Zhao,
  Marimont, Tarroni, Saase, Maier-Hein and Maier-Hein}]{MOODChallenge2020}
\bibinfo{author}{Zimmerer, D.}, \bibinfo{author}{Full, P.M.},
  \bibinfo{author}{Isensee, F.}, \bibinfo{author}{Jager, P.},
  \bibinfo{author}{Adler, T.}, \bibinfo{author}{Petersen, J.},
  \bibinfo{author}{Kohler, G.}, \bibinfo{author}{Ross, T.},
  \bibinfo{author}{Reinke, A.}, \bibinfo{author}{Kascenas, A.},
  \bibinfo{author}{Jensen, B.S.}, \bibinfo{author}{O'Neil, A.Q.},
  \bibinfo{author}{Tan, J.}, \bibinfo{author}{Hou, B.},
  \bibinfo{author}{Batten, J.}, \bibinfo{author}{Qiu, H.},
  \bibinfo{author}{Kainz, B.}, \bibinfo{author}{Shvetsova, N.},
  \bibinfo{author}{Fedulova, I.}, \bibinfo{author}{Dylov, D.V.},
  \bibinfo{author}{Yu, B.}, \bibinfo{author}{Zhai, J.}, \bibinfo{author}{Hu,
  J.}, \bibinfo{author}{Si, R.}, \bibinfo{author}{Zhou, S.},
  \bibinfo{author}{Wang, S.}, \bibinfo{author}{Li, X.}, \bibinfo{author}{Chen,
  X.}, \bibinfo{author}{Zhao, Y.}, \bibinfo{author}{Marimont, S.N.},
  \bibinfo{author}{Tarroni, G.}, \bibinfo{author}{Saase, V.},
  \bibinfo{author}{Maier-Hein, L.}, \bibinfo{author}{Maier-Hein, K.},
  \bibinfo{year}{2022}.
\newblock \bibinfo{title}{{MOOD 2020: A public Benchmark for
  Out-of-Distribution Detection and Localization on medical Images}}.
\newblock \bibinfo{journal}{IEEE Transactions on Medical Imaging} .
\bibitem[{Zimmerer et~al.(2020)Zimmerer, Isensee, Petersen, Kohl and
  Maier-Hein}]{Zimmerer2020Abstract:Auto-encoders}
\bibinfo{author}{Zimmerer, D.}, \bibinfo{author}{Isensee, F.},
  \bibinfo{author}{Petersen, J.}, \bibinfo{author}{Kohl, S.},
  \bibinfo{author}{Maier-Hein, K.}, \bibinfo{year}{2020}.
\newblock \bibinfo{title}{{Abstract: Unsupervised anomaly localization using
  variational auto-encoders}}, in: \bibinfo{booktitle}{Informatik aktuell}.
\bibitem[{Zimmerer et~al.(2019)Zimmerer, Kohl, Petersen, Isensee and
  Maier-Hein}]{zimmerer2019context}
\bibinfo{author}{Zimmerer, D.}, \bibinfo{author}{Kohl, S.},
  \bibinfo{author}{Petersen, J.}, \bibinfo{author}{Isensee, F.},
  \bibinfo{author}{Maier-Hein, K.}, \bibinfo{year}{2019}.
\newblock \bibinfo{title}{Context-encoding variational autoencoder for
  unsupervised anomaly detection}, in: \bibinfo{booktitle}{International
  Conference on Medical Imaging with Deep Learning--Extended Abstract Track}.

\end{thebibliography}

\clearpage

\setcounter{section}{0}
\setcounter{table}{0}
\setcounter{figure}{0}

\begin{center}
\textbf{\large Supplemental Materials. \\ Constrained unsupervised anomaly segmentation.}
\end{center}

\section{On the role of the gradients on VAEs.}
\label{sec:role_gradients}

In this section, we describe the empirical analysis on the gradients role in attention-based anomaly detection using VAEs. To this end, a VAE is trained on normal brain MRI images, and attention maps are extracted for anomalous images. Concretely, we extract Grad-CAMs as defined in Eq.\ref{eq:gradcams}, and non-weighted activation maps (AMs) as following Eq. \ref{eq:am}. A representative case is shown in Figure \ref{fig:am_vs_gradcam} of the main manuscript. Under the explored setting, VAEs Grad-CAMs produce similar attention maps compared to so solely AMs. In particular, we could not find any benefit on gradients weighting other than serving as an scaling factor for attention maps to fall on non-saturated range of values of typically used activation functions, such as sigmoid operation in Eq.\ref{eq:gradcams}. Although Grad-CAMs have been widely used in discriminative models to discern regions of interest in the image using class-specific gradients, its usefulness in generative models such as VAEs seems to be limited.  In this case, the information encoded in the VAE seems closely related to the patterns detected by the convolutional filters in their early layers, without discarding any task-specific information.

\begin{figure}[h!]
    \begin{center}

          \subfloat[\label{fig:ablation_reconstruction_1}]{\includegraphics[width=0.8\linewidth]{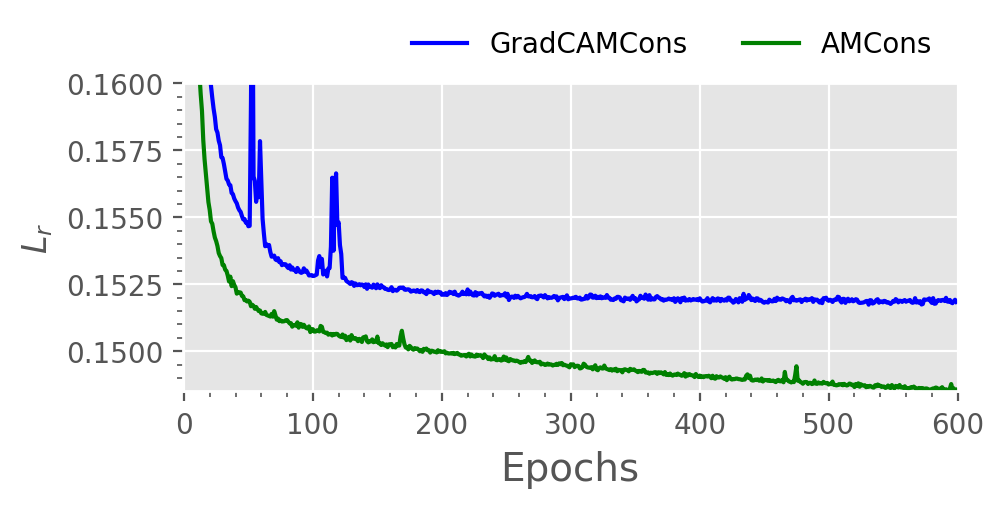}}
          \vspace{-1em}
          \subfloat[\label{fig:ablation_reconstruction_2}]{\includegraphics[width=0.8\linewidth]{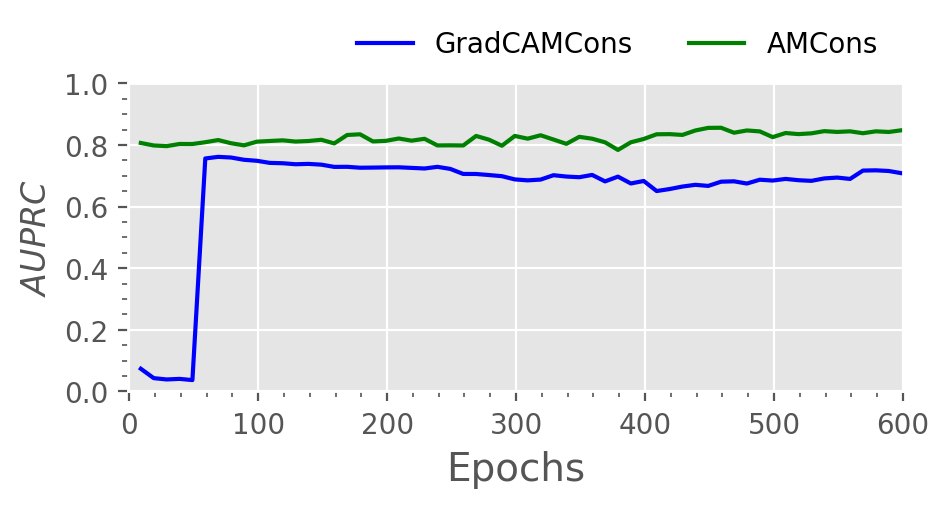}}

        \caption{Study on the gradient influence on image reconstruction. Concretely, we compare grad-CAM based attention constraint (blue) and solely activation map regularization on reconstruction losses (a) and pixel level localization performance (b). Both methods are shown using the best hyperparameters obtained from their respective ablation experiments.}
        \label{fig:ablation_reconstruction}
        \vspace{-2 em}
    \end{center}
\end{figure}

\subsection{Reconstructed images}

\begin{figure}[h!]
\begin{center}
\centering
\includegraphics[width=1\linewidth]{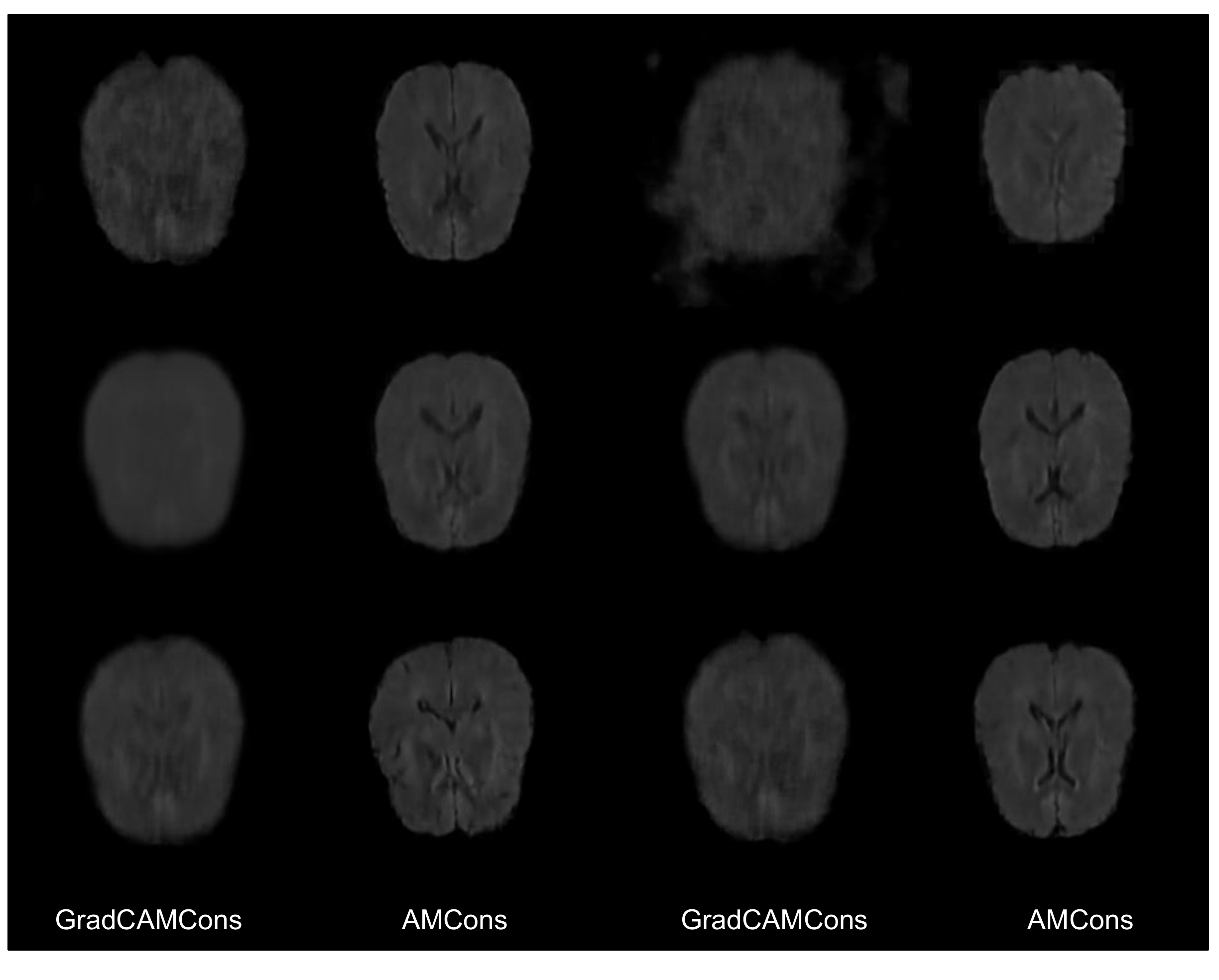} 
\caption{Image reconstruction examples obtained from regularizing gradient-based attention maps (GradCAMCons) and activation maps (AMCons).}
\vspace{-2 em}
\label{fig:reconstructed-images}
\end{center}
\end{figure}

In addition, we also studied the differences of applying constraints on attention maps using gradients (GradCAMCons setting in Eq. \ref{eq:criterion}), or only activation maps \ref{eq:criterion_Lh} in terms of the quality of the reconstructed images. For this purpose, we show in Figure \ref{fig:ablation_reconstruction_1} the learning curves of reconstruction criteria for both methods in their optimal configurations (validated in their respective ablation experiments). In addition, we also show the corresponding results in terms of anomaly localization in Figure \ref{fig:ablation_reconstruction_2}. While the setting based on solely activations maps (AMCons) achieves the best performance, it is also able to bring the lowest reconstruction error. This may be because applying a direct supervision on gradients is too restrictive to optimize the VAE as a whole, compared to the softer criterion of entropy maximization in activation maps we found that the reconstructed images obtained with the GradCAMCons model have lower quality than those provided by the AMCons formulation. Several examples are depicted in Figure \ref{fig:reconstructed-images}.

\vspace{-1 em}
\section{Extraction of brain tissue $\Omega_B$}
\vspace{-1 em}

The AMCons formulation proposed in this work constraints the activation maps of brain tissue to be activated homogeneously, following Eq. \ref{eq:criterion_Lh}. This training procedure requires to extract brain tissue pixels, $\Omega_B$, from background. To do so, we apply an Otsu's threshold to the image to obtain a binary tissue mask. Then, the mask y processed using a morphology closing operation with a disk-shaped structural element of size $5\times5$ pixels. This approach for brain tissue extraction is capable of accurately separating the background from the foreground robustly, due to the observable difference in intensity between the two regions.

\section{Additional ablation experiments.}
\label{sec:additional_ablation}

In the following we present additional ablation experiments that justify the different hyperparameters used in the proposed methods during the experimental stage.

\begin{figure}[h!]
\begin{center}
\centering
\includegraphics[width=1\linewidth]{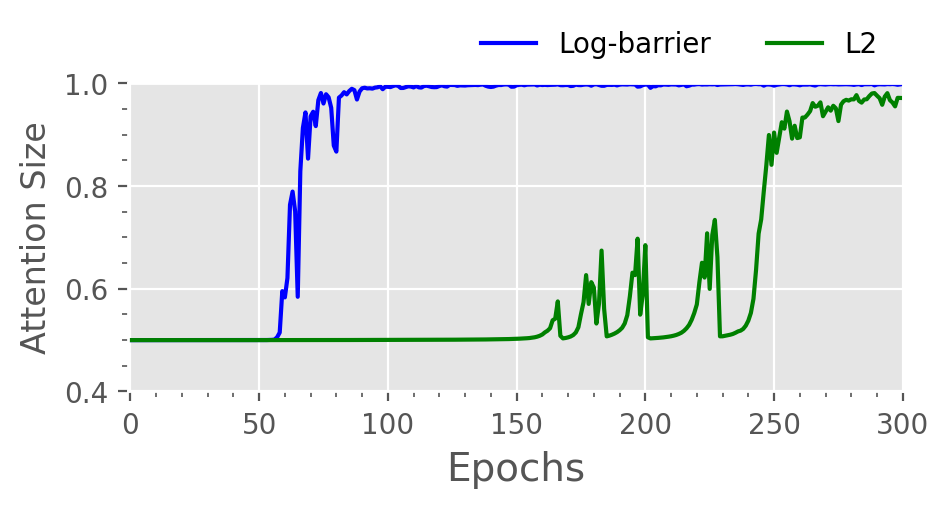} 
\caption{Study on the training dynamics produced by the different constraint criteria, i.e. log-barrier extension and L2 penalty.}
\label{fig:training_dynamics}
\end{center}
\end{figure}

\paragraph{\textbf{On the different approaches to size regularization}} In order to enforce the attention maps to cover the whole normal images during VAE training, \cite{Venkataramanan2020AttentionImages} uses multiple penalties (one per pixel), that forces the activation to be maximum. However, it is well-known in optimization that a penalty does not
act as a barrier near the boundary of the feasible set \cite{boyd2004convex}. In other words, a constraint that is satisfied results in a null penalty and gradient. Therefore, at a given gradient update, there is nothing that
prevents a satisfied constraint from being violated, causing oscillations between competing constraints and ultimately resulting in a potential unstable training. This limitation motivates the methods we propose in this work, which uses a single constraint per image using the average activation. In addition, we advocate for the use of the
log-barrier extension versus penalties due to the strictly positive gradient of the latter becomes higher when a satisfied constraint approaches violation during optimization. The improved dynamics during training are illustrated in Figure \ref{fig:training_dynamics}, which shows how the pixel-level penalty-based method (green line) shows a more unstable convergence.

\vspace{-2 em}
\paragraph{\textbf{On the impact of the reconstruction losses}}We evaluate the effect of including several well-known reconstruction losses in our formulation: SSIM and L$_2$-norm. Table \ref{table:recons_loss} reports the results from these experiments, where we can observe that, while BCE and SSIM reconstruction losses yield the best performances, integrating the L$_2$-norm loss in our formulation degrades the performance of the proposed model.    

\begin{table}[h!]
\centering
\small
\begin{tabular}{|l|c|c|c|}
\hline
       & \textbf{BCE} & \textbf{L2 norm} & \textbf{SSIM} \\ \hline\hline
GradCAMCons w. $\mathcal{L}_S$ & $\mathbf{0.728}$ & $0.612$ & $0.667$ \\ \hline
AMCons w. $\mathcal{L}_H$      & $\mathbf{0.806}$ & $0.714$ & $0.749$ \\ \hline
\end{tabular}
\caption{Ablation study on the reconstruction losses for the proposed approach. The metric presented is the pixel-level AUPRC. Best results in bold.}
\label{table:recons_loss}
\end{table}
\vspace{-1 em}

\paragraph{\textbf{GradCAMCons setting optimization}} To better understand the behaviour of the attention constraints in the proposed using Grad-CAMs and the attention expansion constraint ($\mathcal{L}_S$), we resort to extensive ablation experiments to determine the optimal values of several model hyperparameters: the log-barrier $t$ term, the weights of the attention loss on the training, $\lambda_{s}$ and, finally, the network depth used to compute the CAMs. Firstly, we empirically fix $\lambda_{s}=10^3$, $\beta=1$ and use the first convolutional block output to compute CAMs, to evaluate the impact of our model with $t$ values incldued in $\{1, 10, 25, 50\}$. These results are reported in Table \ref{ablation_t_p}. Please note that all the results reported on the ablation studies are obtained on the validation set.

\begin{table}[h!]
\centering
\footnotesize
\begin{tabular}{|c|ccccc|}
\hline
      & \multicolumn{5}{c|}{t}                                                                                                                                       \\ \hline
      & \multicolumn{1}{c|}{1}            & \multicolumn{1}{c|}{10}           & \multicolumn{1}{c|}{25}           & \multicolumn{1}{c|}{50}           & Scheduler    \\ \hline\hline
AUPRC & \multicolumn{1}{c|}{0.503} & \multicolumn{1}{c|}{$\mathbf{0.728}$} & \multicolumn{1}{c|}{0.670} & \multicolumn{1}{c|}{0.691} & 0.670 \\ \hline
$\lceil$DICE$\rceil$ & \multicolumn{1}{c|}{0.530} & \multicolumn{1}{c|}{$\mathbf{0.693}$} & \multicolumn{1}{c|}{0.649} & \multicolumn{1}{c|}{0.668} & 0.653 \\ \hline
\end{tabular}
\caption{Ablation study on the impact of $t$ in the proposed formulation, where dataset specific AUPRC results are presented. Bold highlights the best performing configuration.}
\label{ablation_t_p}
\end{table}
\vspace{-1 em}

The use of log-barrier extension constraints favour the model optimization and thus the performance on anomaly localization (see Table 3 in the main paper). Nevertheless, this configuration requires to empirically fix more parameters (i.e. $t$ in Eq.4) than the formulation using a $L_2$ penalty. To alleviate this issue, we explore to use a predefined scheduler that incrementally increments the slope $t$ during training. Concretely, the $t$ value is scheduled such that $t=1*1.01^e$, with $e$ being the training epoch. The results presented in Table \ref{ablation_t_p} shows a slight decrease of the results compared with the best configuration ($t=10$). Nevertheless, the obtained results still outperform by a large margin the use of penalty-based methods (see Table 3 in the main paper), as well as other baselines.\\

We now validate the level depth from the encoder used to obtain the CAMs (i.e., network depth $s$ in Section 3.2), with the best configuration from the previous ablation in Table \ref{ablation_t_p}. Results are presented in Table \ref{ablation_cams}, from which we can observe that maximizing the attention in early layers leads to better results than in deeper layers. This could be produced by the better spatial definition of early layers, and the benefits that the proposed constrain produces in its later layers, which receive information from the whole image.\\

\vspace{-1 em}
\begin{table}[h!]
\centering
\footnotesize
\begin{tabular}{|l|c|c|c|c|}
\hline
            & Conv1          & Conv2 & Conv3 & Conv4            \\ \hline\hline
AUPRC       & \textbf{0.728} & 0.639 & 0.579 & 0.105           \\ \hline
$\lceil$DICE$\rceil$  & \textbf{0.693} & 0.632 & 0.583 & 0.196            \\ \hline
\end{tabular}
\caption{Ablation study on network depth to compute CAMs. Dataset specific AUPRC is presented for each possible configuration. Best performance highlighted in bold.}
\vspace{2 em}
\label{ablation_cams}
\end{table}

The experiments presented on the main paper are obtained using the best configuration: $t=10$, $\beta=1$ and $\lambda_{s}=10^3$, with CAMs being obtained form the first convolutional block.

\vspace{-1 em}
\paragraph{\textbf{Using size constraints on AMCons}}

The use of Grad-CAMs on unsupervised anomaly segmentation is supported by the claim that gradients from the latent space allow the discrimination between normal and anomalous regions, as discussed in prior literature (\cite{Venkataramanan2020AttentionImages, Liu2020TowardsAutoencoders}). Nevertheless, we empirically found that in the VAE, gradients are highly correlated with simply the intermediate activations in the encoder, without any discriminating function (see Section 3.4 and Section 1 of Supplemental Material). Therefore, we propose to use only the activation maps in the constrained formulation, in order to not force the gradients during the VAE optimization (AMCons method). In this context, the use of a sigmoid activation (which saturates the activation values) to enforce size supervision that maximizes this activation has certain drawbacks. For example, the value of the activation maps depends on the architecture used, with higher values typically found in deeper layers. A direct result of this could be that, in the absence of gradient-based scaling, the activations are originally in the saturated zone of sigmoid activation. Furthermore, producing an artificial increase in activation values can move the generative model away from its stable configuration, damaging the encoding and reconstruction tasks. For these reasons, in the AMCons configuration, we make use of softmax activation, which normalizes the activations relative to the whole set of pixels, smoothing the applied supervision, while not forcing the activation values to settle around any value. The observer drawbacks of this configuration are confirmed by the empirical results, which are much worse than the proposed entropy constraint, $\mathcal{L}_H$, as shown in Table \ref{ablation_am_ls}.

\begin{table}[h!]
\centering
\footnotesize
\begin{tabular}{|l|cc|}
\hline
      & \multicolumn{1}{c|}{AMCons w. $\mathcal{L}_s$} & AMCons w. $\mathcal{L}_H$ \\ \hline
AUPRC               & \multicolumn{1}{c|}{0.547}    & \textbf{0.728}   \\ \hline
$\lceil$DICE$\rceil$    & \multicolumn{1}{c|}{0.521}    & \textbf{0.693}   \\ \hline
\end{tabular}
\caption{Ablation study on the use of size constraints ($\mathcal{L}_s$) in the activation maps based configuration, AMCons.}
\label{ablation_am_ls}
\end{table}

\vspace{-1 em}
\section{Model complexity.}
\label{sec:model_parameters}

In this section, we compare our formulation to existing approaches in terms of model complexity. Since previous residual-based methods require the generation of normal counterparts from anomalous images, they typically integrate an additional discriminator to create more realistic images, and require to use the trained generative decoder during inference. On the other hand, another interesting property of CAM-based anomaly detection is that it does not require using a decoder during inference stage. As indicated in Table \ref{parameters}, the proposed methods require less computational workload during inference. This phenomenon accentuates using AMCons method, since it does not needs gradients computed from the latent representation, but only intermediate activation maps on the encoder. Moreover, during training, the cost of adding a single constraint is negligible during training, as pointed out in previous literature on constraint optimization (\cite{Kervadec2019Constrained-CNNSegmentation}).

\begin{table}[h!]
\centering
\footnotesize
\resizebox{\linewidth}{!}{
\begin{tabular}{|l|c|c|}
\hline
\multicolumn{1}{|c|}{\textbf{Method}} & \multicolumn{2}{c|}{\textbf{$\sim$Parameters (millions)}}                     \\ \hline
                                      & \multicolumn{1}{c|}{Train} & \multicolumn{1}{c|}{Inference}    \\ \hline\hline
Context VAE (\cite{zimmerer2019context})                            & 15.0     & 15.0    \\ \hline
VAE ((\cite{Baur2019DeepImages,Zimmerer2020Abstract:Auto-encoders}))              & 15.0       & 15.0  \\ \hline
F-anoGAN (\cite{Schlegl2019F-AnoGAN:Networks})                                    & 17.8     & 15.0    \\ \hline
GradCAMCons w. $\mathcal{L}_S$                                                    & 15.0     & 13.3    \\\hline 
AMCons w. $\mathcal{L}_H$                                                         & 15.0     & 3.3     \\ \hline
\end{tabular}
}
\caption{Parameters of the proposed method and best performing baselines during both, training and inference stages.}
\label{parameters}
\end{table}
\vspace{-1 em}

\section{Additional qualitative visualizations}

In the following Figure \ref{fig:qualitative_sm_brats} and Figure \ref{fig:qualitative_sm_ich}, we show complementary examples of the proposed method performance for both datatets: BraTS and Physionet-ICH, respectively.

\newpage

\begin{figure}[h!]
\begin{center}
\vspace{-25 mm}
\centering
\includegraphics[width=1\linewidth]{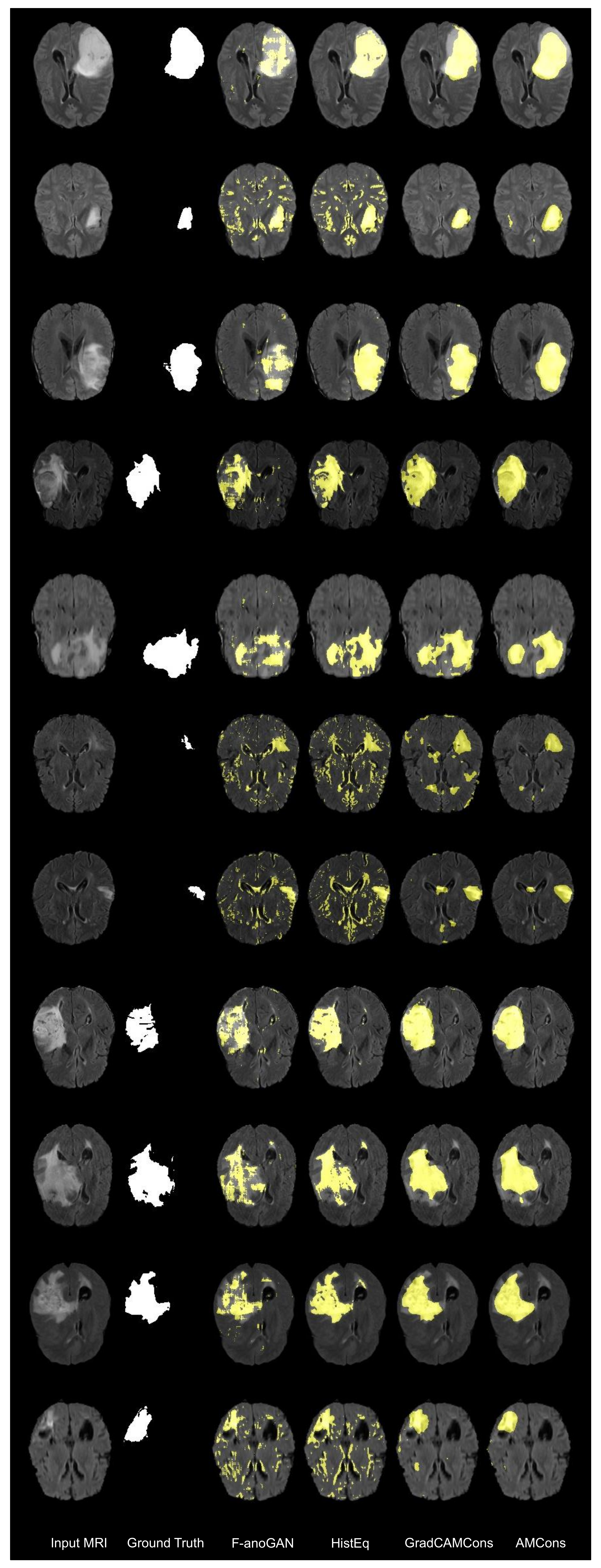} 
\caption{Supplemental qualitative evaluation of the proposed and existing high-performing methods for anomaly localization on BraTS MRI flair volumes.}
\label{fig:qualitative_sm_brats}
\end{center}
\end{figure}

\begin{figure}[htb]
\begin{center}
\vspace{-121 mm}
\centering
\includegraphics[width=1\linewidth]{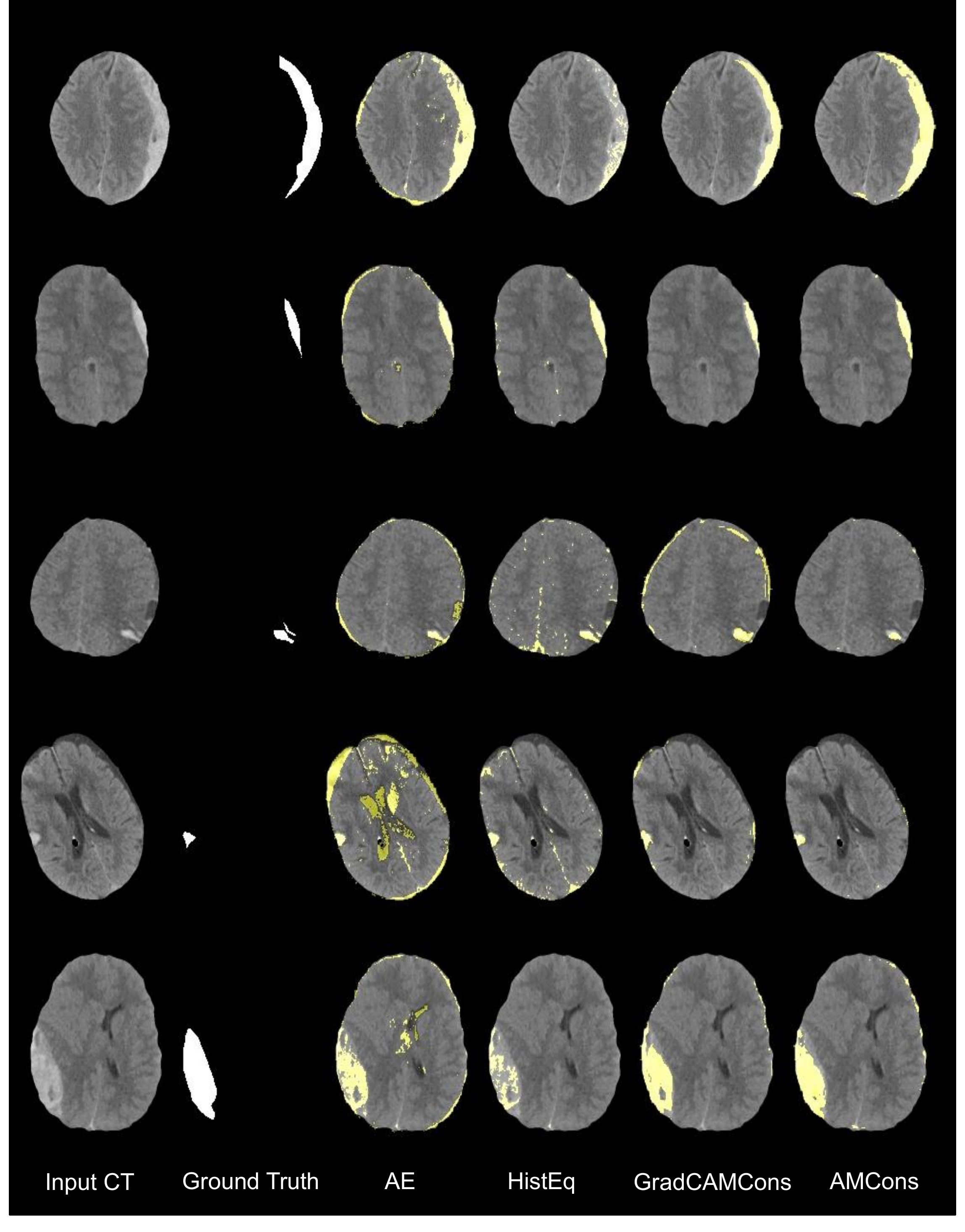} 
\caption{Supplemental qualitative evaluation of the proposed and existing high-performing methods for anomaly localization on Physionet-ICH non-contrast CT images.}
\label{fig:qualitative_sm_ich}
\end{center}
\end{figure}

\end{document}